\documentclass[12pt]{article}

\usepackage[toc,page]{appendix}
\usepackage{slashed}
\usepackage[sumlimits,intlimits,namelimits]{amsmath} 
\usepackage[english]{babel}
\usepackage{graphicx}
\newcommand{\be}{\begin{equation}}
\newcommand{\ee}{\end{equation}}
\newcommand{\ba}{\begin{eqnarray}}
\newcommand{\ea}{\end{eqnarray}}

\newcommand{\ice}[1]{\relax}
\newcommand{\eps}{\varepsilon}

\newcommand{\MSbar}{\overline{\rm MS}}

\usepackage{amsmath}
\usepackage{subfigure}% useful for many pictures
\begin{document}

\thispagestyle{empty}

  \begin{flushright}
SI-HEP-2015-15    \\
QFET-2015-18\\
%Version of \today
  \end{flushright}

\begin{center}

{\bf Inclusive weak decays of heavy hadrons with power 
suppressed terms at NLO}

Thomas Mannel, Alexei A. Pivovarov, Denis Rosenthal

Theoretische Physik 1,Universit\"at Siegen, D-57068 Siegen, Germany

\end{center}

%\preprint{SI-HEP-2015-15}
%\preprint{QFET-2015-18}

\begin{abstract}\noindent
Within the heavy quark expansion techniques for the heavy hadron weak
decays
we analytically compute
the coefficient of the power suppressed dimension five 
chromo-magnetic operator at next-to-leading order of QCD 
perturbation
theory with the full dependence on the final state quark mass.
We present explicit expressions for the total width of inclusive
semileptonic decays including the power suppressed terms
and for a few moments of
decay  differential distributions.
One of the important phenomenological applications of our results 
is precision analysis of the 
decays of bottom mesons to charmed final states and extraction of
the numerical value for the CKM matrix entry $|V_{cb}|$.
\end{abstract}

%\maketitle
%\newpage
%\tableofcontents

\newpage

\section{\label{sec:intro}
Introduction}
Presently the Standard Model of fundamental interactions 
is being thoroughly tested experimentally at colliders,
but no definite signs of New Physics 
have been detected beyond the framework of the Standard Model.
Neither new particles have been explicitly seen nor 
any significant deviations from the Standard Model  
values in the loop sensitive Wilson coefficients
for flavor changing 
observables have been determined in high precision 
data (as a review, see e.g.~\cite{Charles:2015gya}). 
Thus, the Standard Model has successfully passed all tests in 
the areas where it is certainly
valid as a low-energy effective theory. 

However, there is definitely life beyond SM. Some new 
phenomena -- like neutrino masses and 
mixing - can be readily incorporated in a rather straightforward
manner to extensions of SM. The other new effects -- 
like dark matter -- are of cosmological nature and related to still
poorly understood realm of gravity and, strictly speaking, are outside 
the physics of the standard model domain. 
Nevertheless it seems certain 
that the scale of the traditionally expected  
extensions of the standard model -- like 
supersymmetry or extra dimensions -- has definitely
moved from few TeV region to a higher one in 
energies that can make it unreachable at accelerators
in foreseeing future, e.g.~\cite{Forte:2015cia}.
Since the New Physics 
scale moved higher 
the direct observation of new physics phenomena will not probably 
be explicit even at new machines (still one should wait for the results
of the 14~TeV run of LHC!).
In case that nothing will be seen the new
phenomena beyond the standard model (if any at all!)
can only be identified through 
detecting slight discrepancies between
theoretical predictions within the SM
and precision  measurements 
at low energy with available tools.

Accurate theoretical predictions
within the SM are of crucial importance in such a scenario. 
For these predictions to be reliable one first needs
the precise 
numerical values for the key parameters of the SM itself.
The least precisely quantitatively known  
sector of the SM is a quark flavor one where the quark 
Yukawa couplings to the Higgs field 
are not well known numerically. 
In the standard model  
they translate into the mixing
angles between generations gathered in the CKM matrix
and the vacuum expectation 
value of the Higgs field. The latter can be
determined from the leptonic sector.
Note that the flavor sector is also a most promising place 
in investigating the
Higgs mechanism that is definitely of an effective origin and probably
will be modified in future as the presence of a fundamental scalar in
the ``final'' theory 
does not look
convincing.
All in all the flavor physics of quarks  
is the promising place to search for new
physics and should be thoroughly 
studied~(see, e.g.~\cite{Butler:2013kdw,Bevan:2014iga}).

While the quark weak decays are mediated by the charged weak 
currents at 
tree level, which are believed not to have sizable contributions 
of possible 
new physics, their study is of
importance for precise determination of the numerical values
of the CKM matrix elements.
However, obtaining 
solid theoretical predictions for processes with quarks 
at the fundamental level requires the use of genuinely
nonperturbative computational methods like QCD lattice calculations
since eventually one has to make prediction for the experimental
quantities that include hadrons and cannot be described in
perturbation theory of QCD 
due to confinement. 
This is principal part of the problem but there is also a pure
technical part. Even if 
the direct computation in terms of quarks would be relevant to the
world of hadrons that partly can be made possible by choosing proper
observables one will still face the problem of computational 
complexity of 
the calculation with sufficient accuracy that requires a 
rather large
order of perturbation theory. The example is the description of the
process $b\to s\gamma$.

Taking just the parton level of computation for hadronic processes
one makes the technical
part equivalent to that of the leptonic calculations where  
the benchmark level for the technical 
part of the computation is the evaluation of the muon lifetime.
The muon decay is a source for the
determination of the Fermi constant $G_F$ with high 
accuracy from a leptonic sector. First radiative corrections 
have been
computed long time
ago~\cite{Kinoshita:1958ru,Berman:1958ti}. 
To match the precision of the present 
experimental data for muon lifetime,  
the theoretical calculations have to be performed with very high
accuracy.
In this case the calculations are feasible,  
since the purely leptonic decays are well described within
perturbation theory and the expansion parameter 
$\alpha\approx 1/137$ is small.
The latest theoretical result includes the  
second order (NNLO) radiative corrections
in the fine structure
constant expansion~\cite{vanRitbergen:1998yd}
\begin{equation} \label{eq:muon_rate}
\Gamma(\mu \to \nu_\mu e\bar{\nu}_e)
=\frac{G_F^2 m_\mu^5}{192 \pi^3}
\left\{1+\left(\frac{25}{8}-\frac{\pi^2}{2}\right)
\frac{\alpha_r}{\pi}+(6.743+\Delta \Gamma^{had})
\left(\frac{\alpha_r}{\pi}\right)^2\right\} \, .
%\nonumber
\end{equation} 
Here 
$m_\mu$ is the muon mass.
The numerical value for the electron mass $m_e$ is set to zero everywhere
but in the expression for the
expansion parameter
\begin{equation} \label{eq:alphar}
\alpha_r=\alpha+\frac{2\alpha^2}{3\pi}\ln\frac{m_\mu}{m_e}\, .
\nonumber
\end{equation} 
The expressions with account for
nonvanishing electron mass are known. 
The quantity
$\Delta \Gamma^{had}=-0.042\pm 0.002$ is 
the hadronic contribution that is known
with uncertainty of about 5\%. It cannot be computed from first
principles for light quarks and is obtained by integrating the
experimental data for the photon vacuum polarization.
Note that the similar situation emerges with precision analysis one of
the key leptonic observable -- the muon anomalous magnetic moment
$g-2$.
At present the hadronic contributions related to light quarks give the
main uncertainty of theoretical 
prediction~(e.g.~\cite{Pivovarov:2001mw,Kuhn:2003pu}). 
It is a general feature
that quark sector influences even pure leptonic processes if the
required accuracy is high enough, e.g.~\cite{Petrov:2013vka}.

Eq.~(\ref{eq:muon_rate}) 
results in an ${\cal O}$(1ppm) accuracy of the theoretical expression
for the lifetime
that is competitive for precision comparison with  
modern experimental data.
As for the quark sector is concerned
there is a good set of data for $s\to u$ weak 
transitions
that corresponds to $K\to \pi e{\bar {\nu}}_e$ decays at the hadron level, 
but it is hopeless to compute the related rate theoretically at present 
because of strong infrared problems in theoretical treatment of 
reactions with light hadrons.

For heavy hadrons the theoretical treatment of the decays
is however possible 
because the large mass of the heavy quark constitutes a perturbative
scale that is much larger than $\Lambda_{QCD}$. The leading 
logarithmic effects
related to that scale have been discussed long ago~\cite{Shifman:1984wx}.
Later there have been created a framework for 
the possibility for an 
expansion in powers of $\Lambda_{QCD}/m_Q$ where $m_Q$ is the quark 
mass and $\Lambda_{QCD} \sim 500~{\rm MeV}$ is a typical hadronic 
scale~\cite{Georgi:1990um,Neubert:1993mb,Manohar:2000dt}. 
Top quarks do not form mesons 
due to their short lifetime, 
charmed mesons are probably not heavy
enough, rendering the convergence in the inverse mass marginal,
but the case of bottom-meson decays is 
certainly tractable in this way and thus has been
intensively studied.
The technique is applicable to 
$b \to u$ and $b\to c$ transition and both to semileptonic and purely
hadronic inclusive decays. For definiteness, we will stick to
semileptonic $b \to c $ 
decays.

In the present paper
we analytically compute
the coefficient of a power suppressed dimension five 
chromo-magnetic operator at next-to-leading order of QCD 
perturbation
theory with the full dependence on the final state quark mass.
The results of the analogous computation in the massless limit for the
final state quark have been
presented earlier in ref.~\cite{Mannel:2014xza}
Here we present explicit expressions for the total width of inclusive
semileptonic decays and few moments of differential
distributions with full dependence on the final state quark mass.
One of the important phenomenological applications of our results 
is precision analysis of the 
decays of bottom mesons to charmed final states and an extraction of
the numerical value for the CKM matrix entry $|V_{cb}|$.

The paper is organized as follows. 
In the next section we give a
general representation for the decay width of a heavy hadron 
in a form suitable for computation in
QCD. In Sect.~\ref{sec:HQET} we 
give necessary basics of Heavy Quark Effective 
Theory (HQET)
that is a working tool for the present calculation.
In Sect.~\ref{sec:HQE}  we write down 
the Heavy Quark Expansion (HQE) for the decay rate.
The actual computation and results 
are described in Sect.~\ref{sec:calcul}. 
In Appendices we give 
the explicit expressions for our master integrals and some long
analytical expressions for the coefficients of HQE.

\section{\label{sec:rate}
QCD representation for the decay rate}
It is difficult to compute an hadronic decay rate since the
underlying theory of strong interactions -- QCD -- is formulated in
terms of quarks and the hadrons only appear in the strong coupling
regime as bound states.
Therefore one can use either numerical calculation on the lattice 
or find special 
observables for which perturbation theory 
calculation is feasible in some form.
Such observables are inclusive ones since 
the sum over hadronic states
can be related to the sum over the quark-gluon 
states using unitarity of the theory.
In case the initial state is treatable in perturbation theory, 
i.e. it is a leptonic
one as in $e^+e^-$-annihilation into hadrons
or hadronic $\tau$-lepton decays
then the results can be uniquely obtained in perturbation theory. 
In cases when
the initial state is hadronic, i.e. it 
is non-treatable in perturbation theory,
one uses
a factorization idea -- to separate scales 
and compute the short
distance effects in perturbation theory 
while long distance properties are coded in 
hadronic
matrix elements.
The famous example of the latter approach is the analysis of 
deep inelastic scattering of leptons on hadrons. 
The analogue of deep inelastic scattering 
in heavy quark physics is
inclusive decays of heavy hadrons. One can use either fully 
hadronic (non-leptonic)
decays or semileptonic ones. 
The number of experimental observables
in inclusive hadronic decays is however limited to basically 
the total rate of the process. In semileptonic decays the presence of 
leptons in the final states
gives more kinematical flexibility still retaining the rigorous
theoretical description of the process.

The low-energy effective Lagrangian 
${\cal{L}}_{\rm eff}$
for semileptonic $b\to c$
transitions 
is a Fermi four-fermion one 
\begin{equation}\label{eq:Fermi_lagr}
{\cal L}_{\rm eff} = 2\sqrt{2}G_FV_{cb}(\bar{b}_L \gamma_\mu c_L) 
(\bar{\nu}_L \gamma^\mu \ell_L) + h.c.
\end{equation} 
with left-handed fermion fields.
The numerical value for Fermi constant $G_F$ is determined 
from pure leptonic weak processes
and known with high precision.
The mixing angle $V_{cb}$ is the main interest in 
decay measurements with hadronic initial 
states~\cite{Penin:1998wj}.
The precision analysis of such processes is important 
both for the flavor sector and  Higgs mechanism investigations 
in search for new physics.

Using unitarity of the ${\cal S}$-matrix the inclusive decay rate
$B\to X_c\ell\bar{\nu}_\ell$
is obtained from taking the absorptive part of 
the forward matrix element of the transition operator
${\cal T}$~\cite{Bigi:1993fe} that is the 
second order term of the perturbation theory 
expansion in the interaction Lagrangian ${\cal L}_{\rm eff}$,
\begin{equation}\label{eq:trans_operator}
{\cal T} = i\!\!\int \! dx\,    
T\left\{ {\cal L}_{\rm eff} (x)  {\cal L}_{\rm eff} (0) \right\} \, .
\end{equation} 
Note that the transition operator ${\cal T}$
is a non-local functional of the particle fields 
and is given by the integral over all possible scales.
There is no much hope to handle such an operator in QCD 
that includes all scales
as well and no large parameter is available 
in case of the two-point correlator in Eq.~(\ref{eq:trans_operator}). 
However, one can
hope that some transitions or matrix elements are still 
short distance dominated even if it is not a universal
feature of the correlator given in 
Eq.~(\ref{eq:trans_operator})
itself and may depend on external states.
For light hadrons (like kaon) it is definitely not the case and taking
matrix elements cannot help in isolating a short distance dominant
part of the correlator in~\eqref{eq:trans_operator}.
Heavy hadrons have an additional simplification that makes 
the computation of some matrix elements 
possible in perturbation theory by separating the scales 
involved.

The idea is that 
when taking a matrix element over a heavy hadron 
containing a
heavy quark with mass $m_Q\gg \Lambda_{QCD}$
the correlator 
does acquire a large internal 
scale, $m_Q$,  that 
enables scale separation. 
For actual separation of scales one applies
the operator product expansion (OPE)
techniques.
These ideas are formalized through the notion of effective  
theories.
Within the heavy hadron with momentum $p_H$ and
mass $M_H$ the large part
of the momentum is due to a pure kinematical contribution of the
heavy quark $p_H=m_Q v+\Delta$ with $v=p_H/M_H$ 
being the velocity of
the hadron and $\Delta$ is related to the light
degrees of freedom and interactions between them and the 
heavy quark. One can already extract the factor
related to the large quark part of the momentum
explicitly at the level of field variables
when afterwards the matrix element
over a heavy hadron is taken.
The heavy quark field can be separated  
into the fast
oscillating phase and a slow changing field  $h_v(x)$ 
with a typical momentum of order 
$\Delta\sim \Lambda_{QCD}$
\begin{equation}\label{eq:heavy-quark-me-phase}
Q(x)\sim e^{-i(m_Qv) x}h_v(x)\, .
\end{equation}
The velocity $v=p_H/M_H$ is finite in 
the limit of infinitely heavy quarks $m_Q\gg \Lambda_{QCD}$.
This 
program is realized within the effective theory 
for heavy quarks.
In order to make the dependence of the decay width on the
heavy quark mass $m_Q$ explicit and to build up 
an expansion in $\Lambda_{QCD}/m_Q$,  
one matches 
a time-ordered product of full QCD operators in entering to the
transition operator ${\cal T}$ 
onto an expansion in terms of Heavy Quark Effective Theory 
(HQET)~\cite{Mannel:1991mc,Manohar:1997qy}.
Presently the Heavy Quark Expansion in inclusive
semileptonic $b \to c $ 
transitions
provides a level of theoretical precision
in 
the prediction of the total inclusive 
rate for $B \to X_c \ell \bar{\nu}_\ell$ within  
two percent. 
The structure of the HQE
is given by~\cite{Benson:2003kp}
\begin{eqnarray} \label{rate-0}
\Gamma (B \to X_c \ell \bar{\nu}_\ell) =
\Gamma^0 |V_{cb}|^2 \left[ a_0(1 + \frac{\mu_\pi^2}{2m_b^2})
 +  a_2 \frac{\mu_G^2}{2m_b^2}  
 + a_3\frac{\bar{\rho}_3}{m_b^3}  
+ a_4\frac{\bar{\rho}_4}{m_b^4} 
+ {\cal O}\left( \frac{\Lambda_{QCD}^4}{m_b^4}\right)\right]  
\nonumber
\end{eqnarray} 
where 
$\Gamma^0=G_F^2 m_b^5/(192 \pi^3)$ and $m_b$ is the $b$-quark mass.    
The precise definition and the proper choice of the most suitable mass
parameter for the heavy quark field is extensively discussed in 
the literature.
The power suppressed terms are given by the forward matrix 
elements of the local operators of growing dimensionality 
in HQET over the heavy hadron state.
Their numerical values are determined by the corresponding power of
the QCD infrared parameter of $\Lambda_{QCD}$. 
These are nonperturbative quantities either to be computed within
some non-perturbative techniques such as lattice QCD
or to be fitted to
experimental data.
The kinetic energy parameter
$\mu_\pi^2$ is given by the nonrelativistic kinetic energy operator of
the heavy quark within the heavy hadron.
The chromo-magnetic parameter $\mu_G^2$ is given by the matrix element
of the magnetic dipole
operator. These two operators give the leading power suppressed
contribution and were intensively studied.
The higher order power suppressed terms are becoming important
at present as the experimental data improves.
The parameter $\bar{\rho}_3$ describes the contribution of dimension
six operators that are Darwin term and spin-orbit interaction.
The general parameter $\bar{\rho}_4$ is a 
contribution of a rather large number of 
dimension seven operators~\cite{HigherOrders}.
The coefficients $a_i$ are functions of 
the quark and lepton
masses and have a perturbative expansion in the 
strong coupling constant $\alpha_s (m_b)$. 
The leading term coefficient $a_0$ is known analytically
to~${\cal O}\left(\alpha_s^2\right)$ 
precision in the massless limit of the final 
state quark~\cite{vanRitbergen:1999gs}.
At this order 
the mass corrections have been  analytically accounted 
for the total width as an expansion in final fermion mass
in ref.~\cite{Pak:2008qt} 
and for the differential distribution numerically 
in~\cite{Melnikov:2008qs}. 
The coefficient of the 
kinetic energy parameter is linked to the coefficient 
$a_0$ by Lorentz invariance, 
see the explicit analysis 
in~\cite{Becher:2007tk}. 
The NLO correction to the coefficient of the 
chromo-magnetic parameter $a_2$
has been investigated recently 
in~\cite{Alberti:2013kxa} 
where the differential distribution has been
computed and the total decay rate has been then obtained by a process
of numerical
integration over the phase space.
The $\alpha_s$ correction to the chromo-magnetic parameter coefficient
$a_2$
has been analytically computed in ref.~\cite{Mannel:2014xza}
in the massless limit.
Here we give the result with full mass dependence in analytical form.
Our calculation of the coefficient $a_2$ is in fact 
a matching computation 
between QCD and HQET. For this reason
we present some facts about HQET 
relevant for our discussion
in the next section.

\section{\label{sec:HQET}
Basics of HQET}
A heavy quark 
near its mass-shell is described by a field $h_v(x)$ which is
a remnant of the whole QCD fermion field 
$Q(x)$. In fact, it effectively contains only large
components of the Dirac bi-spinor 
that describe the quark and not the antiquark. 
One achieves the separation of the components
by using the projector $P_+=(1+\slashed{v})/2$ where $v$ is the
external velocity that determines the remnant fields  $h_v(x)$
and the whole construction of HQET.
Note that obtaining HQET as the effective theory from QCD 
is very close in spirit to the well known procedure of 
obtaining the nonrelativistic limit of QCD or, earlier, QED. 
The field variables
and Lagrangians are just the same in both 
nonrelativistic QCD and HQET.
The quark velocity $v$ is fixed in the presence of the 
heavy hadron by its
momentum. 
Usually the common choice for the velocity is 
$v=p_H/M_H$. 
The behavior of time and space components of the formal 
Lorentz four-tensors 
differs in HQET.
It is useful to split a four-vector $p^\mu$ in longitudinal and
transverse parts, namely  
$p^\mu=v^\mu (v p)+p^\mu_\perp$. 
The covariant 
derivative of QCD is 
$\pi_\mu =i\partial_\mu+g_sA_\mu $ with the splitting 
$\pi^\mu =v^\mu (v\pi)+\pi^\mu_\perp$. 

The quantity 
$h_v$ is the heavy-quark
field entering the HQET 
Lagrangian~\cite{Mannel:1991mc,Manohar:1997qy}.
The effective Lagrangian of HQET
can be obtained in a concise form at tree level by integrating out the
$P_-$ part of the heavy 
quark field $H_v$, $H_v=P_- H_v$, with the result 
\begin{equation}\label{eq:HQET-Lagr-1}
{\cal L}_{HQET}=
{\bar h}_v(\pi v)h_v
+{\bar h}_v\slashed{\pi}_\perp\frac{1}{2m+\pi v}\slashed{\pi}_\perp
h_v\, .
\end{equation}
Here the first term is just the residual energy of the quark while the
second one describes the effects of the removed (integrated out)
antiquark.
It is non-local that is the price for integrating the antiquark
out. In the limit $m\gg \pi v$ one can expand the second term in 
a series in the inverse large mass and obtain
a local Lagrangian up to a given order in the mass expansion
\begin{equation} \label{eq:HQET-Lagr-2}
{\cal L}_{HQET}=
{\bar h}_v(\pi v)h_v
+{\bar h}_v\slashed{\pi}_\perp\frac{1}{2m}\slashed{\pi}_\perp
{\bar h}_v
-{\bar h}_v\slashed{\pi}_\perp\frac{\pi v}{4m^2}\slashed{\pi}_\perp
h_v 
\, .
\end{equation}
It is inconvenient to have time derivatives in a term that is
formally a correction since then the fields $h_v$
are not correctly canonically
normalized.
Therefore the
redefinition of the fields
is used to remove time derivatives 
\begin{equation}\label{eq:HQET-redefini}
h_v\to \left(1
+\frac{\slashed{\pi}_\perp\slashed{\pi}_\perp}{8m^2}\right) h_v
\end{equation}
and get 
the Lagrangian for the new modes $h_v$ (for which we retain the same
notation though)
in the form
\begin{equation}
{\cal L}_{\rm HQET}={\cal O}_v+\frac{1}{2 m}({\cal O}_\pi
+C_{\rm mag}(\mu){\cal O}_G)
+\frac{1}{2 m^2}{\cal O}_3
+{\cal O}\left(\frac{\Lambda_{QCD}^3}{m^3}\right)
\end{equation}
with
\begin{equation}
C_{\rm mag}(\mu)=1+\frac{\alpha_s(\mu)}{2\pi}
\left\{C_F+C_A\left(1+
\ln\frac{\mu}{m_b}\right)\right\}
\end{equation}
being the coefficient of chromo-magnetic operator ${\cal O}_G$
including the QCD radiative 
correction of the order $\alpha_s$~\cite{Grozin:1997ih}.
For new modes $h_v$ the terms 
of the order $O(1/m_b^2)$ in the Lagrangian 
contain no time derivative~\cite{Manohar:1997qy,Balk:1993ev}.
Here we introduced the notation used below.
The quantity 
${\cal O}_v={\bar h}_vv\pi h_v$ is the leading 
power energy operator
that is independent of the heavy quark mass and spin and gives the
famous spin-flavor symmetry of HQET.
The quantity 
${\cal O}_\pi={\bar h}_v\pi_\perp^2 h_v$ is a kinetic energy
operator and ${\cal O}_G={\bar h}_v
\sigma_{\mu\nu}[\pi_\perp^\mu,\pi_\perp^\nu]/2 
h_v$ is a chromo-magnetic operator. 
They constitute
classical subleading power operators.
Higher terms are given by 
the operator ${\cal O}_3={\bar h}_v[\slashed{\pi}_\perp,
[\slashed{\pi}_\perp,\pi v]] h_v$ that can further be converted 
into a linear combination of the 
Darwin ${\cal O}_D={\bar h}_v[\pi_\perp^\mu,
[\pi_\perp^\mu,\pi v]] h_v$ 
and spin-orbit term 
${\cal O}_{SL}={\bar h}_v\sigma_{\mu\nu}[\pi_\perp^\mu,
[\pi_\perp^\nu,\pi v]] h_v$, ${\cal O}_3=c_D{\cal O}_D+c_{SL}{\cal
  O}_{SL}$
with coefficients $c_D,c_{SL}$ known at the next-to-leading order 
of perturbative expansion in the strong coupling constant.
The discussion of order $1/m_Q^2$ terms in the HQET Lagrangian 
is relevant for our computation
because of the necessity to precisely fix the definition of the
fields $h_v$ entering the heavy quark expansion.

\section{HQE for the width correlator}\label{sec:HQE}
For further convenience we introduce a normalized transition operator 
$\tilde {\cal T}$
through the relation
\begin{equation}
{\rm Im} {\cal T}=
\Gamma^0|V_{cb}|^2\tilde {\cal T}
\, .
\end{equation}
With the use of heavy quark effective theory 
the heavy quark expansion is simply a matching from QCD to HQET 
\begin{equation}\label{eq:HQE-1}
\tilde {\cal T}=
C_0 {\cal O}_0  + C_v\frac{{\cal O}_v}{m_b}
+ C_\pi \frac{{\cal O}_\pi}{2m_b^2}  
+ C_G\frac{{\cal O}_G}{2m_b^2}\, .
\end{equation}
The local operators~${\cal O}_i$
in the 
expansion~(\ref{eq:HQE-1}) are
ordered by their dimensionality
${\cal O}_0 =\bar{h}_v h_v $,
${\cal O}_v =\bar{h}_v v\pi h_v $, ${\cal O}_\pi
=\bar{h}_v\pi_\perp^2 h_v$,
${\cal O}_G=\bar{h}_v
\frac{1}{2}[\slashed{\pi}_\perp,\slashed{\pi}_\perp] 
h_v$.
The coefficients of these operators 
are obtained by matching the relevant matrix elements between 
QCD and HQET.
Note that after taking a matrix element over the hadronic state
(like the $B$-meson) one can use equations of motion for HQET fields $h_v$
to eliminate the operator ${\cal O}_v$.
By the same token there is an operator 
${\cal O}_5=\bar{h}_v (v\pi)^2 h_v $ that is of higher order in 
the large mass
expansion after going on shell using equations of motion of HQET.
Thus, the expansion~(\ref{eq:HQE-1}) is a matching relation
from QCD to HQET with
proper operators up to dimension five with the corresponding 
coefficient functions. 
The coefficients are independent of external states and 
one can take them at will.
We take a heavy quark on shell and gluons
as external states for matching to QCD.

Note that one can use the full QCD fields for the heavy quark
expansion expansion as well. However the choice of the proper basis of
operators is not so  straightforward as in HQET.
Still
it is convenient to choose the local operator  
$\bar{b}\slashed{v} b$ defined in full QCD as a leading term of 
heavy quark expansion~\cite{Manohar:1993qn}.
Indeed, the current $\bar{b}\gamma_\mu b$  is 
conserved and its forward matrix element with
hadronic states is 
absolutely normalized.
For implementing this setup
one needs an expansion (matching) of 
a full QCD local operator $\bar{b}\slashed{v} b$ in
HQE through HQET operators. The expansion reads
\begin{equation}\label{eq:local-bvb}
\bar{b}\slashed{v} b
={\cal O}_0 - {\tilde C}_\pi\frac{{\cal O}_\pi}{2m_b^2}
+ {\tilde C}_G\frac{{\cal O}_G}{2m_b^2}
+O(\Lambda_{QCD}^3/m_b^3)
\end{equation}
up to necessary order in the strong coupling 
$\alpha_s$. The  coefficient of the 
leading power operator ${\cal O}_0$ has no radiative 
corrections and the kinetic operator has
the coefficient related to the leading one due to Lorentz
(reparameterization)
invariance.

Substituting the expansion~(\ref{eq:local-bvb})
into eq.~(\ref{eq:HQE-1}) one obtains after using the equation
of motion for the operator ${\cal O}_v$  in the forward 
matrix elements
\begin{eqnarray}\label{eq:HQE}
\tilde{\cal{T}}=C_0 \left\{\bar{b}\slashed{v} b  
 -\frac{{\cal O}_\pi}{2m_b^2}\right\}
+ \left\{-C_v C_{\rm mag}(\mu)+C_G - {\tilde C}_GC_0\right\}
\frac{ {\cal O}_G}{2m_b^2}.
\end{eqnarray}  
Note that for phenomenological applications 
the numerical value for the chromo-magnetic moment parameter  
$\mu_G^2 $, related to the forward matrix element of the operator
${\cal O}_G$, is usually 
taken from the mass splitting between the pseudoscalar and 
vector ground-state mesons.
The mass difference of bottom mesons
$m_{B^*}^2-m_{B}^2=\Delta m_B^2 =0.49~{\rm GeV}^2$ 
is given by 
\begin{equation} \label{delM}
\frac{1}{2 M_B }C_{\rm mag}(\mu)\langle B(p_B)| {\cal O}_G(\mu)|B(p_B)\rangle=
\frac{3}{4}\Delta m_B^2
\end{equation} 
(up to higher order $1/m_Q$ corrections)
where we use the relativistic normalization of states. 
Therefore the coefficient in front of the renormalization group 
invariant combination
$C_{\rm mag}(\mu) {\cal O}_G(\mu)$ can be useful.
In such normalization one gets 
after 
taking the forward matrix element of the expansion 
in Eq.~(\ref{eq:HQE}) the representation
\begin{eqnarray}\label{HQEfin}
\Gamma(B\to X_c{\bar \nu}_\ell\ell)
=\Gamma^0|V_{cb}|^2\left\{
C_0 \left(1+\frac{\mu_\pi^2}{2 m_b^2}\right)
+ \left(-C_v+\frac{C_G - {\tilde C}_G C_0}{C_{\rm mag}}\right)
\frac{3\Delta m_B^2}{8m_b^2}\right\} \, . 
\end{eqnarray} 

\section{\label{sec:calcul}
Description of the calculation and results 
}
\subsection{\label{sec:calcul-descr}
Generalities and techniques
}
The matching procedure consists in computing
matrix elements with partonic states (on-shell
quarks and gluons)
at both sides of the expansion~(\ref{eq:HQE-1}).
The coefficient function $C_0$
of the dimension three operator $\bar{h}_v h_v$
determines the total width of the heavy quark and at the same
time the leading contribution to the width of a bottom hadron
with HQE technique. 
At NLO the calculation of the 
transition operator $\tilde{\cal T}$ in~(\ref{eq:trans_operator})
requires to consider three-loop diagrams with external heavy quark
lines on shell.
The leading order 
result is well known and requires the
calculation of the two-loop Feynman integrals of the simplest topology
-- the sunset type ones~\cite{we-annals}.
\begin{figure}[h!]
%\begin{center}
\centering 
\includegraphics[width=0.35\textwidth]{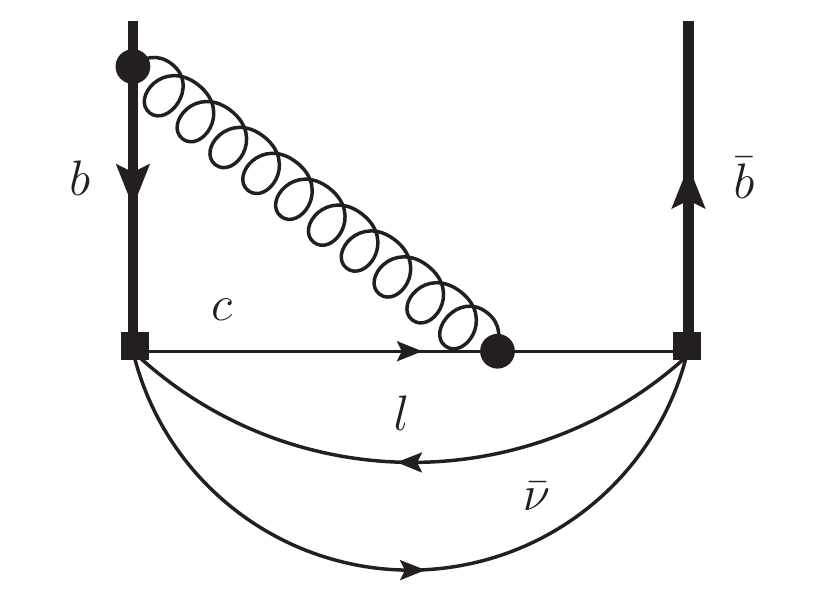}
\qquad
\includegraphics[width=0.35\textwidth]{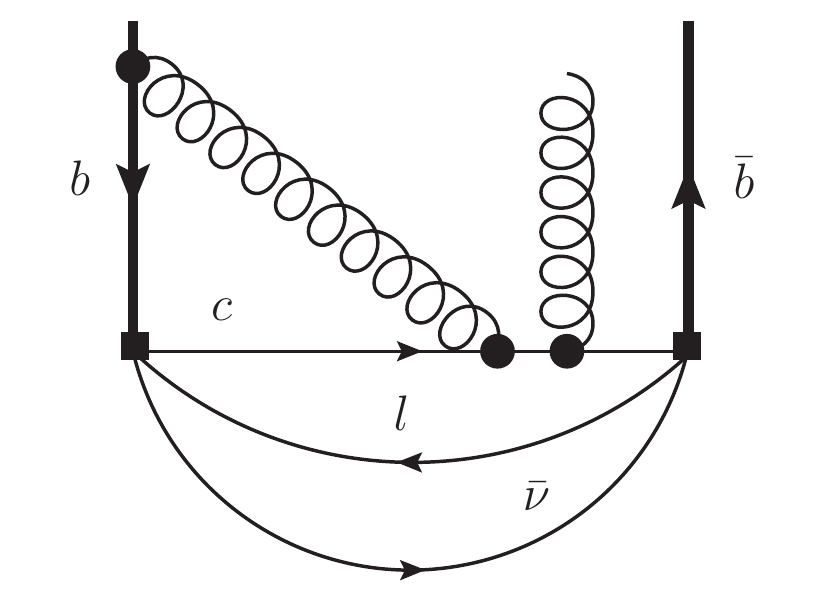}
%\end{center}
%\vspace{-0.5cm}
\caption{\label{fig:diags-1}
Perturbation theory diagrams for the 
matching computation at NLO level, 
(left) - partonic type, right - power correction 
type: insertion of an external gluon}
\end{figure}
At the NLO level 
one needs the on-shell three-loop integrals with massive 
lines due to the massive $c$-quark.
In Fig.~\ref{fig:diags-1} we show some typical three-loop diagrams
both for the partonic part and power corrections 
of the decay rate. 

The computation has been performed in dimensional regularization used for
both ultraviolet and infrared singularities. We used 
the systems of symbolic
manipulations REDUCE~\cite{reduce} and Mathematica~\cite{Mathe}
with original codes written for the calculation. 
The package FeynCalc~\cite{FeynCalc} 
is used for manipulating Dirac matrices and four
vectors under Mathematica.
The reduction to master integrals
has been done within the integration by parts 
technique~\cite{Tkachov:1981wb}. 
The original codes have been used for most
of the diagrams and then the program LiteRed~\cite{Lee:2013mka} 
has been used for checking
and further application to complicated diagrams. 
The master integrals have been computed directly and then checked
with the program HypExp~\cite{Huber:2007dx}.
The renormalization is performed on-shell by the multiplication of the
bare (direct from diagrams) results by the on-shell
renormalization constant 
$Z_2^{OS}$
\begin{equation} \label{eq:Z2OS}
Z_2^{OS}=1-C_F\frac{\alpha_s(\mu)}{4\pi}\left(\frac{3}{\epsilon}
+3\ln\left(\frac{\mu^2}{m_b^2}\right)+4\right) \, .
\end{equation}
It is convenient to fix the normalization point to the $b$-quark mass 
$\mu=m_b$ in the practical computation. The $\mu$-dependence can be
easily restored from the knowledge of anomalous dimensions.  

We present and discuss the obtained results below.

\subsection{The leading power coefficient $C_0$: 
partonic width}
By using the described 
methods we reproduce the known result for the heavy quark width which
is given by the contribution of the leading operator 
${\cal O}_0$. The coefficient $C_0$ is  
\begin{equation}\label{eq:C0}
C_0=C_0^{LO}+C_F\frac{\alpha_s}{\pi}C_0^{NLO}
\end{equation}  
where the LO contribution reads
\begin{equation}\label{eq:C0-LO}
C_0^{LO}=1 -8 r-12 r^2\ln(r)+8 r^3- r^4 
\end{equation}  
and the NLO  contribution reads
\ba\label{eq:LO-C0-NLO}
C_0^{NLO}&=&(1-r^2)\left\{\left(\frac{25}{8}
-\frac{239}{6}r+\frac{25}{8}
  r^2\right)
+\left(-\frac{17}{6}+\frac{32}{3}r-\frac{17}{6} r^2\right) \ln(1-r)\right\}
\nonumber \\
&&+\left(-10-45r+\frac{2}{3} r^2-\frac{17}{6} r^3\right) 
r\ln(r)+\left(-18-\frac{r^2}{2}\right)r^2 \ln ^2(r)
\nonumber \\
&&+\left(2+60 r^2+2 r^4\right) \ln (1-r) \ln (r)
+\left(1 +16 r^2+ r^4\right) (3\text{Li}_2(r)-\pi^2/2) 
\nonumber \\
&&
+16 r^{3/2} (1+r)\left(\pi^2-4
\left(\text{Li}_2(\sqrt{r})-\text{Li}_2(-\sqrt{r})\right)+
2\ln(r)\ln\frac{1+\sqrt{r}}{1-\sqrt{r}}
\right)
\ea  
with $r=m_c^2/m_b^2$. Here $\text{Li}_2(r)$ is polylogarithm,
$\text{Li}_2(r)=\sum_nr^n/n^2$. The combination 
\ba
r^{1/2}\left(\pi^2-4
\left\{\text{Li}_2(\sqrt{r})-\text{Li}_2(-\sqrt{r})\right\}+
2\ln(r)\ln\frac{1+\sqrt{r}}{1-\sqrt{r}}
\right)
\ea
is a part of one master integral in the computation and 
it always appears in thios form.
It contains a specific odd contribution $r^{1/2}\pi^2$ while the rest 
is in fact formally even in $m_c$.
The analytical expression at NLO in Eq.~(\ref{eq:LO-C0-NLO})
has 
been first given by Nir~\cite{Nir:1989rm}.

The behavior near the border of 
the decay phase space ($r\sim 1)$ of the NLO correction
\be
C_0^{NLO}(r\to 1)=\frac{3}{10}(1-r)^5+O((1-r)^6)
\ee
is similar to that of the LO which is 
\be
C_0^{LO}(r\to 1)=\frac{2}{5}(1-r)^5+O((1-r)^6)
\, .
\ee

A typical feature of the result at  
next-to-leading order  
is the presence of odd powers of the charm
quark mass like 
$r^{3/2}$. Of course, it does not mean 
that there is a symmetry $m_c\to -m_c$.
At the small mass limit $r\to 0$
only the simplest term of such structure 
$\pi^2r^{3/2}$ survives with a rather large coefficient. 

We define the bottom quark mass being a pole one because it is 
convenient for computing the relevant matrix elements in QCD 
with on-shell quark states.
The definition of charmed quark mass can be either the pole scheme 
or $\MSbar$-scheme one. 
The relation between the two definitions up to necessary order
is 
\be
m_c^{pol}=m_c^{\MSbar}(\mu)
\left( 1+C_F\frac{\alpha_s}{4\pi}\left(3\ln\frac{\mu^2}{m_c^2}
+4\right)\right)\, .
\ee
The numerical value for the charmed quark mass is
best known in the $\MSbar$-scheme~\cite{Allison:2008xk,PDG}.
It is rather small and cannot be
be perturbatively cast into the pole mass scheme
with any reliable control over uncertainties due to  
convergence of perturbation series 
expansion~\cite{Krasnikov:1995is}.
The numerical value for the bottom quark mass has been discussed in
the literature for a long time and many estimates are available.
Also there is a extensive discussion which particular scheme of
defining the quark mass parameter which is the most 
suitable for this 
particular observable~\cite{Penin:1998zh,Bigi:1994re}.

\begin{figure}[h!]
\centering 
\includegraphics[width=0.621\textwidth]{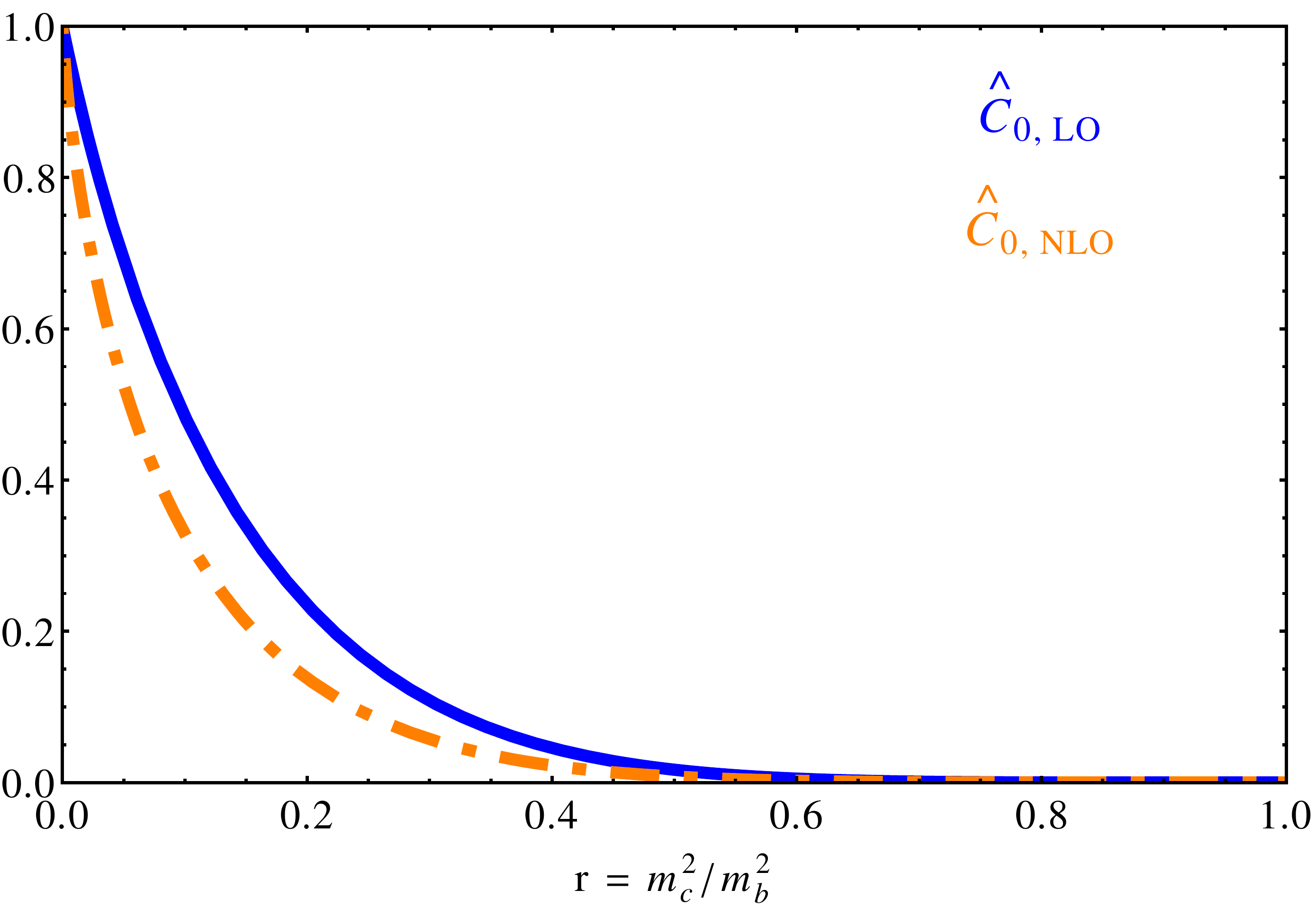}
\caption{\label{fig:c0-pole}
The mass dependence of the coefficient $C_0(r)$ 
in a pole mass scheme
for $m_c$: solid line - LO, dashed-dotted line - NLO normalized to 
$C_0^{NLO}(0)$: ${\hat C}_0^{NLO}(r)=C_0^{NLO}(r)/C_0^{NLO}(0)$,
${\hat C}_0^{LO}(r)=C_0^{LO}(r)/C_0^{LO}(0)$ .}
\end{figure}

In Fig.~\ref{fig:c0-pole} we give the plot of the coefficient
$C_0^{LO}(r)$ and also the normalized next-to-leading coefficient
${\hat C}_0^{NLO}(r)$  in the
pole mass scheme for $m_c$.

\begin{figure}[h]
\begin{center}
\includegraphics[width=0.621\textwidth]{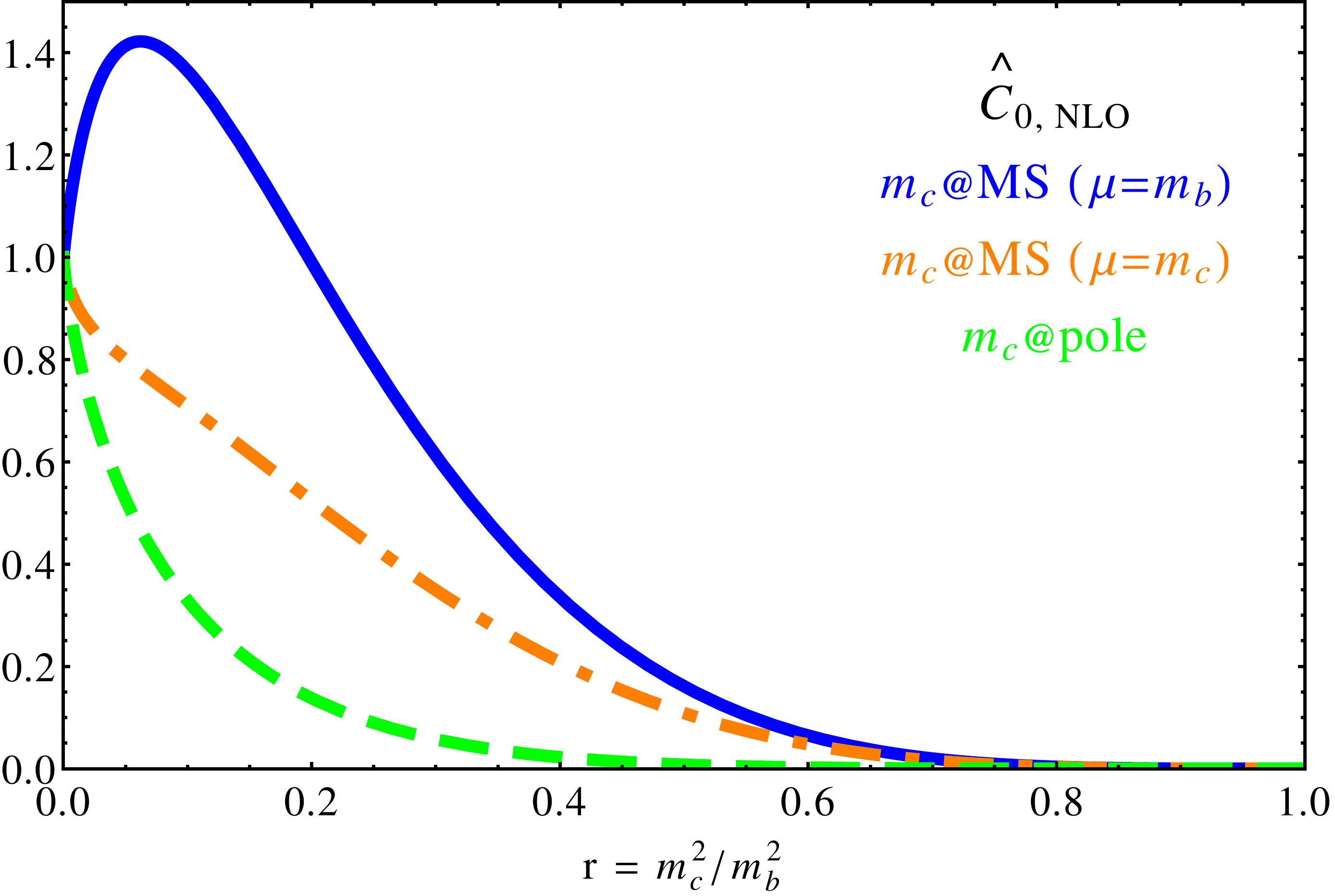}
\end{center}
\caption{\label{fig:c0-MS}
Mass dependence of the coefficient 
$C_0^{NLO}(r)$ in pole and $\MSbar$ schemes
with $\mu=m_b$ and $\mu=m_c$ .}
\end{figure}

In Fig.~\ref{fig:c0-MS} we give the plot of the mass dependence of the
coefficient $C_0^{NLO}(r)$ in different mass schemes for $m_c$.

In the small mass limit for the charmed quark 
one finds
\ba\label{eq:c0-NLO-small}
C_0^{NLO}(r)|_{r\to 0}=\left(\frac{25}{8}-\frac{\pi ^2}{2}\right)
-2 r (6 \ln (r)+17)+16 \pi ^2 r^{3/2}
+ O\left(r^2\ln^2 r\right)
\, .
%\nonumber 
%\\
%&&+r^2
%   \left(-18 \log ^2(r)+18 \log (r)-8 \pi
%     ^2-\frac{273}{2}\right)+O\left(r^{5/2}\right)
\ea
We have computed the results for the coefficient 
$C_0$ in  massless limit,  $C_0(0)$,
independently that serves partly as a
check of our full mass calculation.

The relative magnitude of the 
NLO contribution at a typical value of mass
ratio
$r=0.07$ is
\be
C_0(0.07)=0.60 - 0.78 C_F\frac{\alpha_s}{\pi}
=0.6(1-C_F\frac{\alpha_s}{\pi}1.31)
\ee
while 
in massless limit it is 
\be
C_0(0)=1-C_F\frac{\alpha_s}{\pi}1.8
\, .
\ee
The numerical value for the bottom quark mass 
$m_b$ is important for phenomenological applications and discussed in
the literature (see, e.g.~\cite{Penin:1998zh}).
The dependence on the charm quark mass 
is essential but still it follows mainly the
pattern of that at leading order.
This similarity supports the idea of ref.~\cite{Mannel:2014xza} 
that the computation in massless limit
can be useful for physical applications 
as the normalization and the extrapolation with the leading order 
massive result can be a reasonable approximation for the mass
dependence at NLO. We will see how it works or does not work for other
coefficients later.

\subsection{The $vD$-operator coefficient $C_v$}
Here we present 
the result for the coefficient $C_v$ which is an auxiliary
quantity in our approach since the operator is reexpressed through the
other contributions at the level of matrix elements. 
The coefficient $C_v$ is singled out by taking the matrix element
between $b$-quarks on shell and one gluon with vanishing momentum and
longitudinal polarization, i.e. the gluon field is chosen on the form 
$A_\mu=v_\mu(v A)$. Here $A_\mu$ is a matrix in color space 
$A_\mu=A_\mu^at^a$.
The result for the  coefficient $C_v$
\be
C_v=C_v^{LO}+C_F\frac{\alpha_s}{\pi}C_v^{NLO}
\ee
reads
\ba
C_v^{LO}&=&5-24 r-12 r^2 \ln(r)+24 r^2-8 r^3+3 r^4 \, .
\nonumber
\ea

%&=&\frac{1}{m_b^4}\frac{d}{d m_b}
%\left(m_b^5 C_0^{LO}(m_c^2/m_b^2)\right)
%\ea
%

\begin{figure}
\centering
\includegraphics[width=0.621\textwidth]{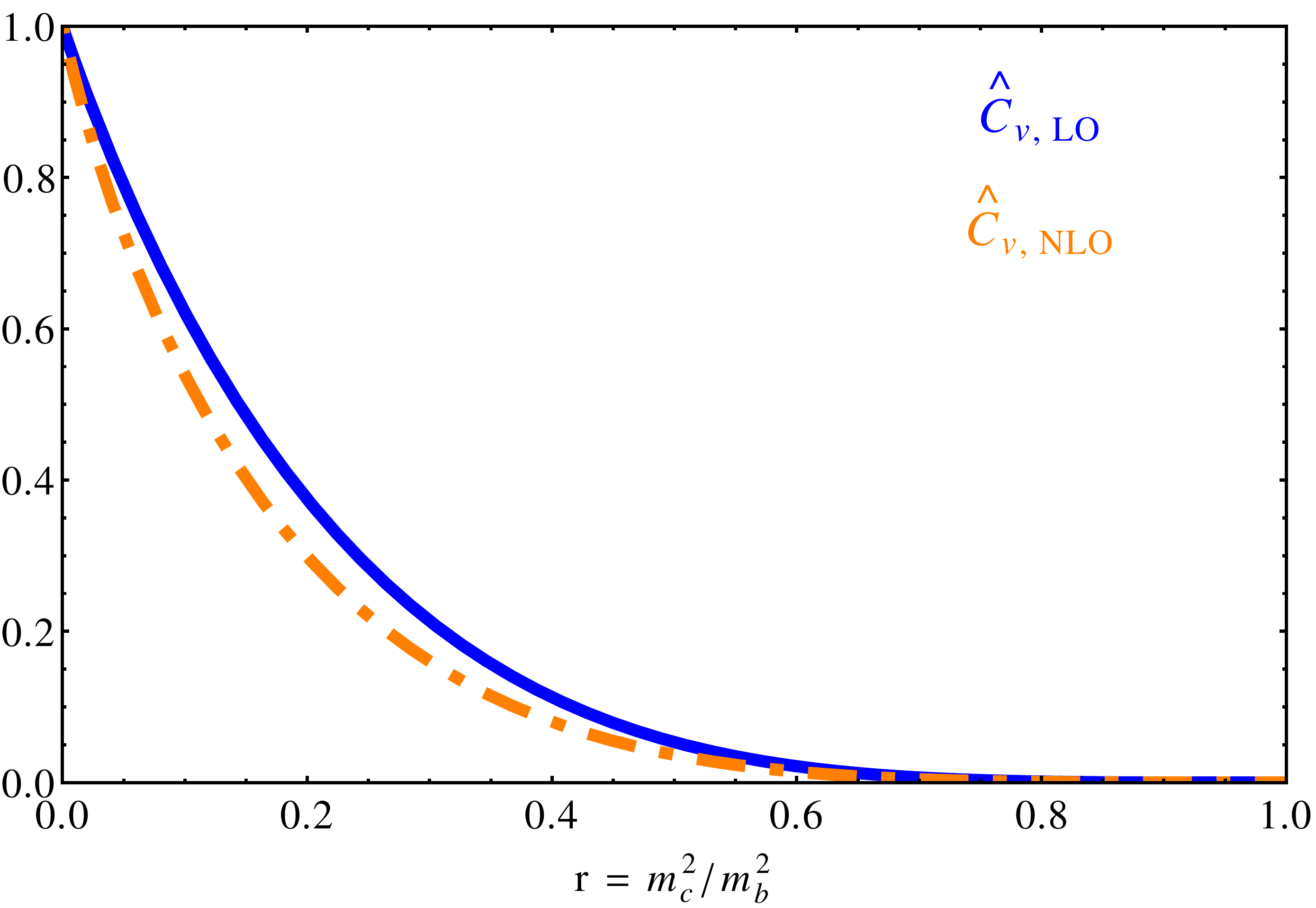}
\caption{\label{fig:cv}
Mass dependence of the coefficient $C_v(r)$ }
\end{figure}

In Fig.~\ref{fig:cv} we plot the charmed quark 
mass dependence of $C_v$.

For the NLO part $C_v^{NLO}$ we give explicitly 
only the expression for the small $r$ expansion. 
The structure of the whole contribution 
is very similar to
that of $C_0^{NLO}$. The expression 
is rather long and given in Appendix~\ref{App:AppendixB}.
The small mass expansion reads
\begin{equation}\label{eq:cv_NLO}
C_v^{NLO}=\left(-\frac{25}{24}-\frac{\pi ^2}{2}\right)
+48 r-8 \pi ^2 r^{3/2}+O\left(r^2\ln^2 r\right)
\, .
%+r^2 \left(6 \log ^2(r)+42
%   \log (r)+8 \pi ^2+\frac{171}{2}\right)+O\left(r^{5/2}\right)
\end{equation}
The leading term of the expression coincides with the 
independent computation in the massless limit done 
in~\cite{Mannel:2014xza} 
\begin{equation}\label{eq:cv_massless}
C_v|_{r=0}=5
+C_F\frac{\alpha_s}{\pi}\left\{
-\frac{25}{24}-\frac{\pi^2}{2}
\right\}\, .
\end{equation}
As for the mass dependence of the coefficient $C_v$, 
for the typical value of $r=0.07$
one finds 
\ba
C_v(0.07)=3.6 -3.8 C_F\frac{\alpha_s}{\pi}=
3.6\left(1 - C_F\frac{\alpha_s}{\pi}1.1\right)
\ea
while in the massless limit one has
\ba
C_v(0)=5(1- C_F\frac{\alpha_s}{\pi}1.2)\, .
\ea
One sees again a rather reasonable accuracy for the mass dependence
extrapolation
at NLO.

The coefficient $C_v$ has no $C_A$ color structure, it contains 
only the $C_F$ Casimir invariant.
This property matches
the possibility to compute this coefficient using a small momentum
expansion near the quark mass shell, $p=mv+k$. 
Still, an explicit cancellation of the
contribution proportional 
to the color structure $C_A$ and cancellation
of poles with the same
renormalization constant~$Z_2^{OS}$ shown in Eq.~(\ref{eq:Z2OS})
is a powerful
check of the final result.

The large $m_c$ behavior at the border of phase space is
\be
C_v^{NLO}(r\to 1)=-3(1-r)^4+O((1-r)^5)
\ee
and 
\be
C_v^{LO}(r\to 1)=4(1-r)^4+O((1-r)^5)
\, .
\ee

\subsection{The coefficient  
$C_G-C_0{\tilde{C}}_G\equiv C_G^r$:\\
chromo-magnetic operator}
\begin{figure}[h!t]
\begin{center}
\includegraphics[width=0.621\textwidth]{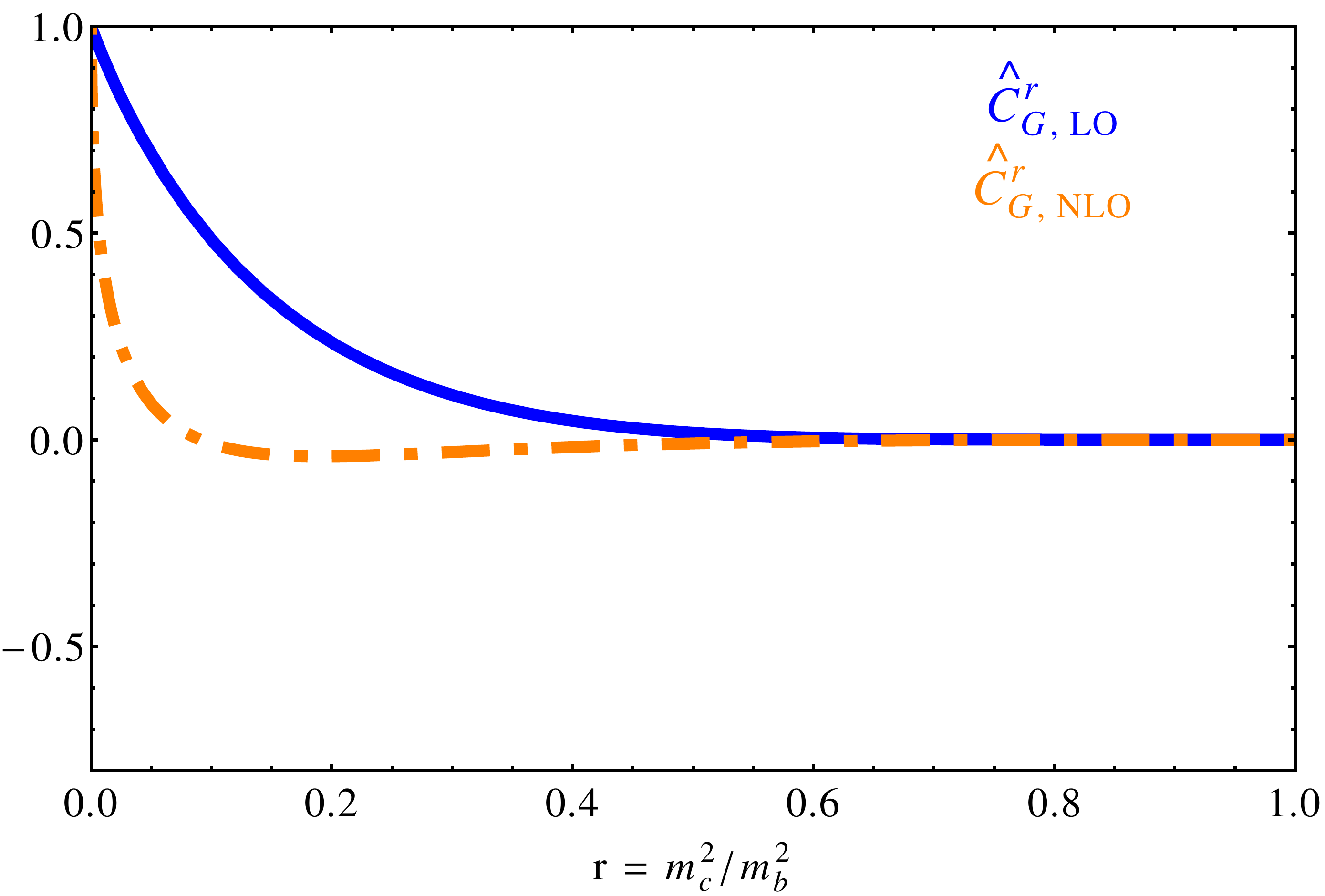}
\end{center}
\caption{\label{fig:cG-c0cG}
Mass dependence of 
the coefficient $C_G^r=C_G-C_0{\tilde{C}}_G$ with $\mu=m_b$}
\end{figure}

For the chromo-magnetic operator coefficient 
we directly compute the difference 
between contributions to the width
correlator in Eq.~(\ref{eq:HQE-1})
and the local $\bar{b}\slashed{v}b$ operator in 
Eq.~(\ref{eq:local-bvb}) multiplied by the leading power 
coefficient
$C_0(r)$, $C_G^r=C_G - C_0{\tilde{C}}_G$.  We write this coefficient 
as leading order term and radiative correction in the form
\be
C_G^{r}(r)=C_G^{r,LO}(r)
+\frac{\alpha_s}{\pi}
\left\{C_A C_G^{r,NLO,A}(r)+C_F C_G^{r,NLO,F}(r)
\right\}
\ee
where the NLO coefficient is separated into 
two color structures with $C_A$ and $C_F$ color group 
invariants.
In Fig.~\ref{fig:cG-c0cG} we present the plot of the
mass dependence for the coefficient of
the chromo-magnetic operator for QCD with $C_A=3$ and 
$C_F=4/3$.
One sees that the mass dependence of 
$C_G^r$ at NLO is much sharper than in
previous cases. This is unexpected and makes the conjecture about
a uniform phase space suppression for the coefficients
less accurate. 
The explicit leading order expression reads
\be
C_G^{r,LO}=2 -16r -24 r^2\ln(r)+16 r^3 -2 r^4 = 2 C_0^{LO}
\, .
\ee
The NLO coefficients with full mass dependence are
too long,
whereas the expanded results are
\begin{eqnarray}  \label{eq:CGr-NLO}
C_G^{r,NLO,A}&=&-\frac{8 \pi^2 \sqrt{r}}{3}
+r \left(\ln ^2(r)-25  \ln (r)+\frac{2 \pi^2}{3}-25\right)
-\frac{\pi^2}{9}+\frac{49}{18}
\, ,
\nonumber \\
C_G^{r,NLO,F}&=&\frac{32 \pi^2 \sqrt{r}}{3}
+r \left(-4 \ln ^2(r)+68 \ln (r)-4 \pi ^2+21\right)
-\frac{7 \pi ^2}{9}-\frac{47}{36} \, .
\end{eqnarray}

At the border of phase space we obtain
\be
C_G^{r,NLO}(r\to 1)=C_F(1-r)^4
+\frac{1}{5}\left[2 C_F - 3 C_A \right](1-r)^5+O((1-r)^6)
\ee
and 
\be
C_G^{r,LO}(r\to 1)=\frac{4}{5}(1-r)^5+O((1-r)^6)\, .
\ee
In the massless limit the $C_G^{r}$ coefficient is given by 
\be
C_G^{r}(0) = 2+\frac{\alpha_s}{\pi}
\left\{
C_A\left(\frac{49}{18}-\frac{\pi^2}{9}\right)+
C_F\left(-\frac{47}{36}-\frac{7\pi^2}{9}\right) 
\right\}
\, .
\ee
This result has been independently 
determined by the 
direct computation using the technology developed for
the massless case.

\subsection{Coefficient 
$C_{\bar{\mu}_G^2}=-C_v+C_G^r/C_{\rm mag}$: \\
the matrix element of 
$C_{\rm mag}{\cal {O}}_G$
}
This coefficient is the final result after the use 
of equations of motion.
We prefer to give the coefficient in front of 
the renormalizaton group invariant combination
that enters the HQET Lagrangian.
This combination also determines the 
mass splitting in the ground state multiplets due to spin orientation.

Thus, the final coefficient of the 
matrix element of the 
chromo-magnetic operator with account of equation of motion after
taking hadronic matrix elements
reads
\begin{equation}
C_{{\bar\mu}_G^2}(r)=-C_v(r)+\frac{C_G^r(r)}{C_{\rm mag}(\mu)} \, .
\end{equation}
This is a coefficient in front of the matrix element 
of the renormalization invariant
combination $C_{\rm mag}(\mu){\cal O}_G(\mu)$.

\begin{figure}[h!]
%\begin{center}
\centering
\includegraphics[width=0.4\textwidth]{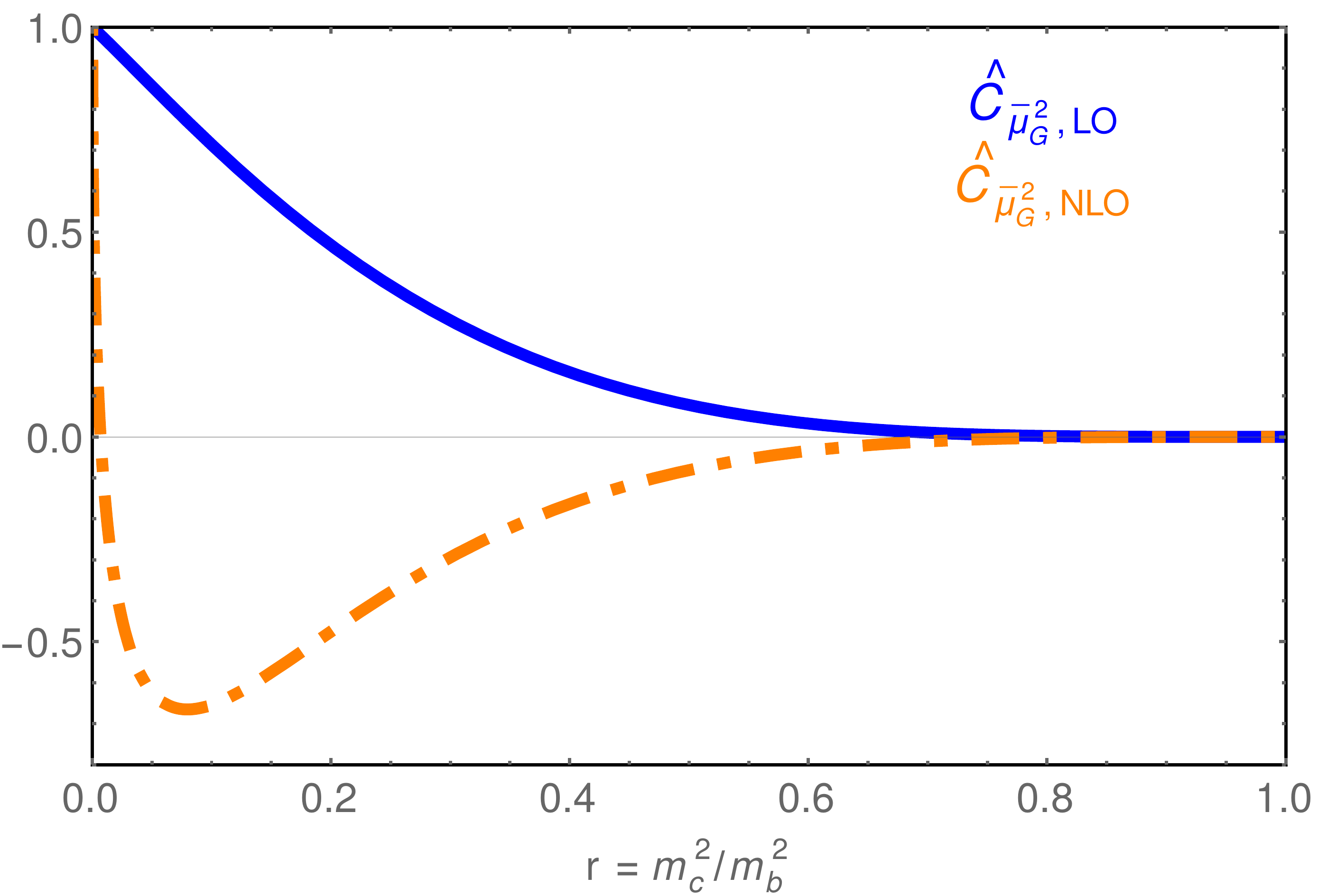}
\qquad
\includegraphics[width=0.4\textwidth]{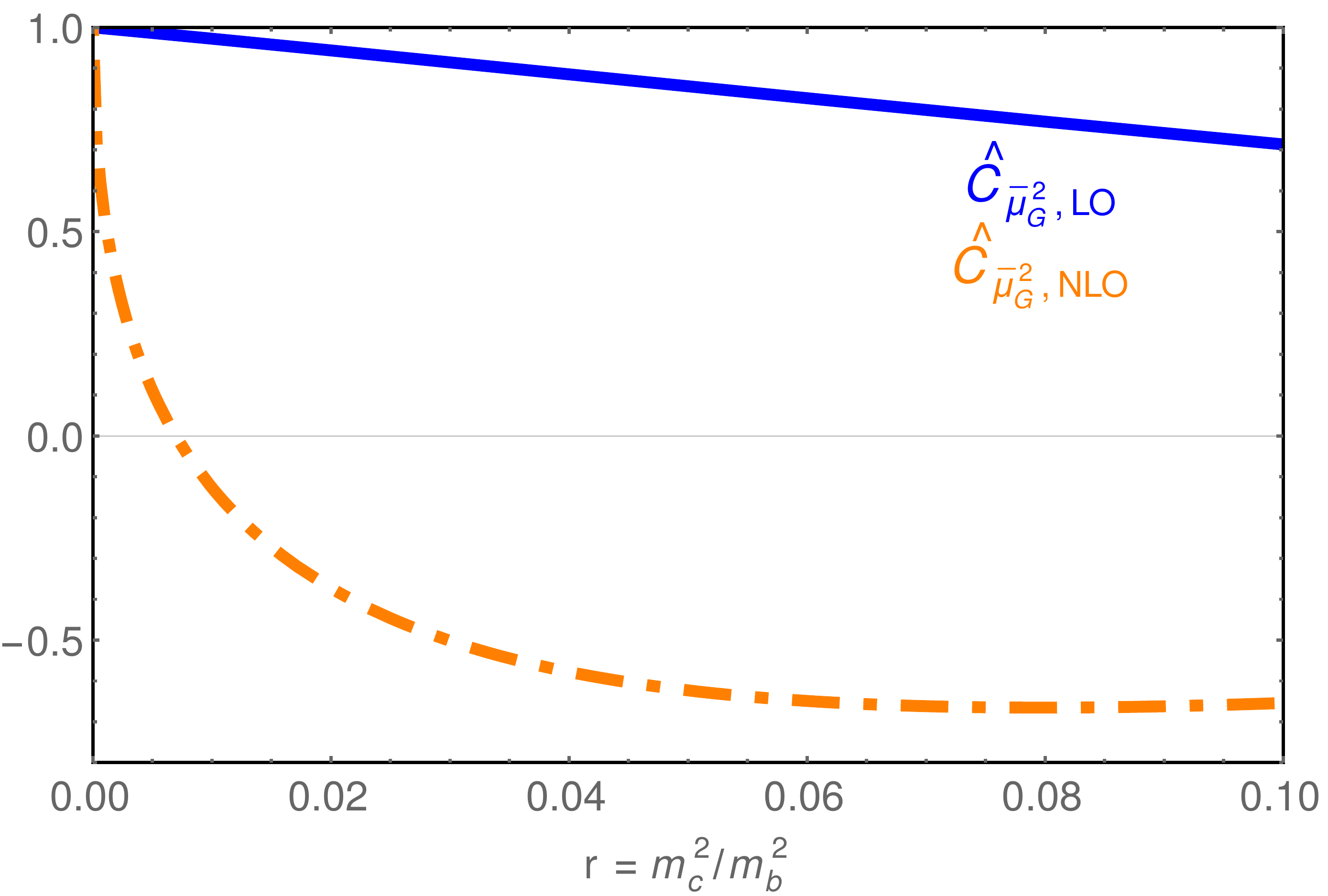}
%\end{center}
%\vskip -2cm
\caption{\label{fig:cGfin}
The mass dependence of the 
coefficient $C_{\bar{\mu}_G^2}(r)$ at LO and NLO in 
the pole mass scheme. Color blind: $C_A=3$,
 $C_F=4/3$. Left panel -- the whole phase space, 
right panel -- zoomed image of the small mass region  
$0<r<0.1$}
\end{figure}

In Fig.~\ref{fig:cGfin} we plot the mass dependence of this final
coefficient. 

Writing 
again the decomposition of the whole coefficient 
in $\alpha_s$ order
\begin{equation}
C_{\bar{\mu}_G^2}(r) = C_{\bar{\mu}_G^2}^{LO}+\frac{\alpha_s}{\pi}
\left\{
C_A C_{\bar{\mu}_G^2}(r)^{NLO,A}+C_F C_{\bar{\mu}_G^2}(r)^{NLO,F}
\right\}
\end{equation}
we obtain at the leading order the well known result  
\begin{eqnarray}  \label{eq:Cfin-LO}
C_{\bar{\mu}_G^2}^{LO} = 
-3 + 8r -24 r^2-12 r^2 \ln (r)+24 r^3 -5 r^4
\, .
\end{eqnarray} 
The whole expressions are given in Appendix~\ref{App:AppendixC}.
Here we present the new result at NLO as 
a small $r$ expansion only
\begin{eqnarray}  \label{eq:Cfin-NLO}
C_{\bar{\mu}_G^2}^{NLO,A}&=&%\left(
\frac{2 \pi ^2 r}{3}-17 r-\frac{8 \pi ^2
    \sqrt{r}}{3}
+r \log ^2(r)-25 r \log
   (r)-\frac{\pi ^2}{9}+\frac{31}{18}\, ,
%\right) 
\nonumber \\
C_{\bar{\mu}_G^2}^{NLO,F}&=&-4 \pi ^2 r-19 r+\frac{32 \pi
   ^2 \sqrt{r}}{3}-4 r \log ^2(r)+68 r \log (r)-\frac{5 \pi
   ^2}{18}-\frac{91}{72}\, .
\end{eqnarray} 
The color blind expansion for QCD ($C_A=3$, $C_F=4/3$)
reads
\ba
C_{\bar{\mu}_G^2}^{NLO}&=&\left(
\frac{94}{27}-\frac{19 \pi ^2}{27}\right)
+\frac{56 \pi  ^2 \sqrt{r}}{9}\nonumber \\
&&+\frac{1}{3} r
   \left(-7 \ln ^2(r)+47 \ln (r)-10 \pi ^2-229\right)
+\frac{280}{27} \pi ^2
   r^{3/2}\nonumber \\
&&+\frac{1}{81} r^2 \left(-1251 \ln ^2(r)-1917 \ln (r)-216 \pi
   ^2-5750\right)+O\left(r^{5/2}\right)\, .
\ea
The very large contribution of the $\sqrt{r}$ term leads to a 
very fast change of the coefficient $C_{\bar{\mu}_G^2}^{NLO}$
from its massless
limit value with an increase of
the charm quark mass. Numerically one finds
\ba
C_{\bar{\mu}_G^2}^{NLO}
=-3.46+61.41 \sqrt{r}+ r \left(-2.3 \ln ^2(r)
+15.7 \ln(r)-109.2\right)+O\left(r^{3/2}\right)\, .
\ea

In the massless limit the new result is
\begin{eqnarray}  \label{Cfin}
C_{\bar{\mu}_G^2}=-3+\frac{\alpha_s}{\pi}
\left\{
C_A\left(\frac{31}{18}-\frac{\pi^2}{9}\right)-
C_F\left(\frac{91}{72}+\frac{5\pi^2}{18}\right) 
\right\} . \nonumber 
\end{eqnarray} 
Note that the $C_F$ part of this coefficient 
differs from the result given in ref.~\cite{Mannel:2014xza}.
The difference is given by $2 C_0^{NLO}$ and it emerged because 
in~\cite{Mannel:2014xza} only the leading order of $C_0$
coefficient 
was used for subtracting the contribution of the local $\bar
b\slashed{v} b$
operator.

The $\mu$ dependence of  the prefactor of 
${\cal O}_G$ in Eq.~(\ref{eq:HQE}) matches the leading order 
anomalous dimension of chromo-magnetic 
operator~\cite{Grozin:1997ih}, 
such that $C_{\bar{\mu}_G^2}$ is  $\mu$ independent. 

The end-of-spectrum behavior reads
\be
C_{\bar{\mu}_G^2}^{NLO}(r\to 1)=
4C_F(1-r)^4+\left[\frac{3}{10}C_F-C_A\right](1-r)^5
+O((1-r)^6)
\ee
for NLO and 
\be
C_{\bar{\mu}_G^2}^{LO}(r\to 1)=-4(1-r)^4+O((1-r)^6)
\ee
for the leading order contribution.

The mass parameter of the heavy quark  $m_b$ 
is chosen to be the pole mass 
which is a proper formal parameter for perturbative 
computations in HQET (see discussion in~\cite{Benson:2003kp}).
After having obtained the results of perturbation theory 
computation for the coefficients of HQE, one is free to change
this parameter to any other~\cite{Bigi:1994re}.

\section{Discussion of the results}

\subsection{The total width}
The radiative corrections are of reasonable magnitude and are well
under control for the numerical values of the coupling constant for 
$\mu\sim 2-4~{\rm GeV}$ (for the numerical value see, 
e.g.~\cite{Korner:2000xk}. 
This provides a clean application of the
results to phenomenology.
The final quark mass dependence is remarkable.
It is very fast for small $m_c$ therefore 
the decays into light quarks~$u$ for bottom mesons and 
$d$~for
charmed mesons should be treated with care.

The coefficients of HQE have been also calculated in 
ref.~\cite{Gambino} where the analytical computation has been
performed for the hadronic tensor and the 
final integration over the phase
space has been done numerically. Such a setup has advantages for
direct comparison with experimental data since the experimental cuts
in the phase space can be readily introduced.

We can make a literal comparison 
with the results of~\cite{Gambino} for the total width.
Our result in the format of ref.~\cite{Gambino} is 
\be\label{eq:gambino-final}
\Gamma=\Gamma_0^m
\left((1 - 1.7776 \frac{\alpha_s}{\pi})(1-\frac{\mu_\pi^2}{2 m_b^2})
-(1.9449 + 2.4235  \frac{\alpha_s}{\pi})\frac{\mu_G^2}{m_b^2}\right)
\ee
for $r=0.0625$ that literally coincides with 
the results of ref.~\cite{Gambino}.

For phenomenological applications and comparison with experiment 
within our approach one can compute 
moments of the differential distribution
(see, e.g.~\cite{Falk:1995me}).
It is straightforward to compute almost any moment in the invariant
lepton pair mass, lepton pair energy or  invariant mass of the
hadronic system. We present few such moments below.

\subsection{Moments of differential distribution}
Note that our computation is organized such that 
it allows for computation of certain moments of differential
distribution. We can build up moments over the 
leptonic pair invariant mass squared $q^2$,
($q=p_\ell+p_\nu$) and the partonic
invariant mass squared $(p-q)^2$, $p$ is the momentum 
of the bottom quark and $p=m_b v$.
It is possible because we have the leptonic part 
and the partonic parts separately 
in an intermediate representation of
computed diagrams -- one can compute
the moments in $q^2$ or/and in $(p-q)^2$.
The total lepton energy moments (the moments 
in the variable $pq$) are
just the linear combinations of those two sets.
We present 
the analytical results for few moments at small $r$ expansion for
brevity.
The analytical expression for the total 
width is given for further comparison with the moments.
It reads
\ba
\Gamma/\Gamma_0&=&
1-8r + C_F\frac{\alpha_s}{4\pi}\left(
\frac{25}{2}-2\pi^2- 8 r \left(6 \ln (r)+17\right)\right)+
\nonumber \\
&&
\frac{\bar{\mu}_G^2}{2m_b^2}\left\{-3+8r+
\frac{\alpha_s}{4\pi}\times\right.
\nonumber \\
&&\left.
\left(C_A \left(\frac{2 \pi ^2 r}{3}-17 r-\frac{8 \pi ^2
    \sqrt{r}}{3}
+r \ln ^2(r)-25 r \ln(r)
-\frac{\pi ^2}{9}+\frac{31}{18}\right)
\right.\right.
\nonumber \\
&&\left.\left.
+C_F \left(-4 \pi ^2 r-19 r+\frac{32 \pi
   ^2 \sqrt{r}}{3}-4 r \ln ^2(r)+68 r \ln (r)-\frac{5 \pi
   ^2}{18}-\frac{91}{72}\right)
\right)\right\} 
\, .
\ea

The normalized $q^2$ 
moments of the total width with $C_{\bar\mu_G^2}$ 
coefficient are given below.
For convenience they are normalized to unity at leading order of
power, small mass, and perturbative expansions. 
The normalization can be
obtained independently.
Indeed,
the $x=q^2/m_b^2$ distribution in massless limit at LO is given by 
\ba
\frac{1}{\Gamma_0}\frac{d\Gamma}{dx}=2(1-x)(1+2x)
\, .
\ea
The normalization factors for the 
moments $n=1-3$
are then 
$N(M^q_n)=\{3/10,2/15,1/14\}$. 
For example,
\ba
N(M^q_1)=\int_0^1 2(1-x)(1+2x) x dx=3/10\, .
\ea

First moment $(q^2/m_b^2)^1$ is 
\ba
&&M_1^q=1-15 r+C_F\frac{\alpha_s}{4\pi}\left(
13-2\pi^2
-r \left(90 \ln (r)+\frac{10 \pi ^2}{9}+355\right)\right)+
\nonumber \\
&&\frac{\bar{\mu}_G^2}{2m_b^2}
\left(-\frac{25}{3}+\frac{\alpha_s}{\pi} 
\left(C_A  \left(-\frac{80 \pi ^2 \sqrt{r}}{9} -
\frac{25 \pi ^2}{27}
+\frac{260}{27}\right)+C_F  \left(\frac{320 \pi ^2 
\sqrt{r}}{9}
+\frac{65
   \pi ^2}{54}-\frac{763}{36}\right)\right)
\right)
\, .
\nonumber 
\ea
Second moment $(q^2/m_b^2)^2$ is 
\ba
&&M_2^q=1-24 r + C_F\frac{\alpha_s}{4\pi}\left(
\frac{604}{45}-2\pi^2
+r \left(-144 \ln (r)-2 \pi ^2-\frac{6813}{10}\right)\right)+
\nonumber \\
&&\frac{\bar{\mu}_G^2}{2m_b^2}
\left(-15+\frac{\alpha_s}{\pi} 
\left(C_A \left(-20 \pi ^2 \sqrt{r}-\frac{17 \pi
   ^2}{12}+\frac{541}{40}\right)
+C_F \left(80 \pi ^2 \sqrt{r}+\frac{7 \pi
   ^2}{3}-41\right)\right)\right)
\, .
\ea
Third moment $(q^2/m_b^2)^3$ is
\ba
&&M_3^q=1-35r +C_F\frac{\alpha_s}{4\pi}\left(
\frac{1243}{90}-2\pi^2 
+r \left(-210 \log (r)-\frac{14 \pi ^2}{5}-\frac{20195}{18}      
\right)\right)+
\nonumber \\
&&\frac{\bar{\mu}_G^2}{2m_b^2}
\left(-23+\frac{\alpha_s}{\pi}\left(C_A  
\left(-\frac{112 \pi ^2 \sqrt{r}}{3}-\frac{35 \pi
   ^2}{18}+\frac{27217}{1620}\right)\right.\right.\nonumber\\
&&\left.\left.\qquad\quad+C_F\left(\frac{448 \pi ^2
   \sqrt{r}}{3}+\frac{67 \pi
   ^2}{18}-\frac{1088429}{16200}\right)\right)\right)
\, .
\nonumber
\ea

The $q^2$ moments are very stable and hardly change with $n$
besides the total normalization. Usually one argues that 
radiative  corrections
should increase or decreases depending on the momentum flow through
the diagram -- we see no simple explanation for the change of
radiative corrections.

The moments in partonic variable $(p-q)^2-m_c^2$ are defined
through the relation
\ba
M_n^H=\int\frac{((p-q)^2-m_c^2)^n}{m_b^{2n}}
\frac{d\Gamma}{\Gamma_0}
\ea
and have been considered in~\cite{Falk:1995me}. 
They are given below for $n=1-3$ analytically 
within small $r$ expansion. 

First moment $((p-q)^2-m_c^2)^1$ is
\ba
&&M_1^H= C_F\frac{\alpha_s}{\pi} 
\left(\frac{71 r}{24}+\frac{3}{2} r \log (r)+\frac{91}{600}\right)+
\nonumber \\
&&\frac{\bar{\mu}_G^2}{2m_b^2}
\left(\frac{\alpha_s}{\pi} \left(
C_A \left(-\frac{611 r}{108}-\frac{22}{9} r \log (r)
-\frac{29}{180}\right)\right.\right.
\nonumber \\
&&\left.\left.
+C_F  \left(-\frac{73 \pi ^2
   r}{36}+\frac{457 r}{108}-\frac{67}{36} r \log (r)-\frac{\pi ^2}{4}+\frac{77}{45}\right)\right)-\frac{3 r}{2}+\frac{1}{2}\right)
\, .
\ea
Second moment $((p-q)^2-m_c^2)^2$ 
\ba
&&M_2^H= C_F\frac{\alpha_s}{\pi}\left(\frac{5}{432}
-\frac{137 r}{600}\right)
\nonumber \\
&&+\frac{\bar{\mu}_G^2}{2m_b^2}
\frac{\alpha_s}{\pi} \left(C_A \left(r\left(\frac{\ln r}{18}
+\frac{163}{1080}\right)+\frac{1}{72}\right)
%\right.
%\nonumber \\
%&&\left.
+C_F \left(r\left(\frac{25   \ln r}{18}+\frac{3703}{1080}\right)
+\frac{347}{3600}\right)\right)
\, .
\ea
Third moment $((p-q)^2-m_c^2)^3$ is
\ba
&&M_3^H=C_F\frac{\alpha_s}{\pi}\left(\frac{377}{176400}
-\frac{119 r}{3600}\right)
%\nonumber \\
%&&
+\frac{\bar{\mu}_G^2}{2m_b^2}
\frac{\alpha_s}{\pi}\left(C_A  \frac{43}{16200}
+C_F \frac{11537}{1587600}\right)
\, .
\nonumber
\ea
This set is such that moments vanish at
leading order. Therefore one cannot discuss the relative magnitude of 
radiative corrections.
Our results for $n=1-2$ coincide with those of 
ref.~\cite{Falk:1995me}).

We also compute the relevant moments numerically with 
full mass
dependence for a typical value of the mass ratio.
For the $q^2$ moments up to third order 
we obtain with $r=0.0625$
\ba
%\hat{\mathcal{M}}_{q^2}^{(0)}
M^q_0&=&(1 -1.7776 \frac{\alpha_s}{\pi})
%\Big(1-\frac{\mu_\pi^2}{2m_b^2}\Big)
-3.8898 (1-0.9206
\frac{\alpha_s}{\pi}) \frac{\bar\mu_G^2}{2m_b^2}
\, ,
\nonumber \\
%\hat{\mathcal{M}}_{q^2}^{(1)}
M^q_1&=&(1 -1.6500 \frac{\alpha_s}{\pi})
%\Big(1-\frac{\mu_\pi^2}{2m_b^2}\Big)
-8.9901 (1-0.6834
\frac{\alpha_s}{\pi}) 
\frac{\bar\mu_G^2}{2m_b^2}
\, ,
\nonumber \\
%\hat{\mathcal{M}}_{q^2}^{(2)}
M^q_2&=&(1 -1.5575 \frac{\alpha_s}{\pi})
%\Big(1-\frac{\mu_\pi^2}{2m_b^2}\Big)
-14.394 (1-0.5578
\frac{\alpha_s}{\pi}) 
\frac{\bar\mu_G^2}{2m_b^2}
\, ,
\nonumber \\
%\hat{\mathcal{M}}_{q^2}^{(3)}
M^q_3&=&(1 -1.4847 \frac{\alpha_s}{\pi})
%\Big(1-\frac{\mu_\pi^2}{2m_b^2}\Big)
-19.997 (1-0.4666
\frac{\alpha_s}{\pi}) 
\frac{\bar\mu_G^2}{2m_b^2}
\, .
\ea
Numerically 
for partonic moments in $H=(p-q)^2-m_c^2$ with $r=0.0625$
one obtains
\ba
%\mathcal{M}_{H}^{(1)}
M_1^H&=&0.0569 \frac{\alpha_s}{\pi}
%\Big(1-\frac{\mu_\pi^2}{2m_b^2}\Big)
+0.397  (1-2.304
\frac{\alpha_s}{\pi}) \frac{\bar\mu_G^2}{2m_b^2}
\, ,
\nonumber \\
%{\mathcal{M}}_{H}^{(2)}
M_2^H&=&0.00575  \frac{\alpha_s}{\pi}
%\Big(1-\frac{\mu_\pi^2}{2m_b^2}\Big)
+0.0554 \frac{\alpha_s}{\pi} 
\frac{\bar\mu_G^2}{2m_b^2}
\, ,
\nonumber \\
%{\mathcal{M}}_{H}^{(3)}
M_3^H
&=&0.00114 \frac{\alpha_s}{\pi}
%\Big(1-\frac{\mu_\pi^2}{2m_b^2}\Big)
+0.00694 \frac{\alpha_s}{\pi} 
\frac{\bar\mu_G^2}{2m_b^2}
\, ,
\ea
where ${\bar{\mu}_G^2}=C_{\rm mag}(\mu)\mu_G^2(\mu)$.

It is also possible to compute the moments of the lepton energy 
spectrum that is of interest 
from the experimental point of view. However, here a few more
technical 
problems arise. On the 
one hand the whole set up of the analytical calculation has to be
modified, 
since leptonic tensor 
has to be taken as a differential distribution rather than fully 
integrated over the lepton phase space. 
On the other hand there is the question of how to deal with $\gamma_5$ 
in dimensional regularization. 
For the cases we discussed here we always have a situation 
when there is an even (in
fact two) number of $\gamma_5$-matrices within 
the trace over Dirac matrices both in leptonic and hadronic
parts, so we simply and consistently use anticommuting $\gamma_5$. 
However, in the calculation of the 
moments of the charged-lepton energy one has also 
consider an odd number
of $\gamma_5$-matrices in the traces, which causes an additional 
complication of the calculation. 
Nevertheless, with the technology developed here, 
these problems 
can be tackled and we plan 
to present a calculation of lepton-energy moments 
in a separate publication. 

\subsection{\label{sec:pheno-look}
Phenomelogical outlook
}

This paper has been devoted to the description of the technical 
aspects of the calculation of the perturbative 
QCD corrections for subleading powers in the $1/m$ expansion. 
Aside from more theoretical consideration, 
such as the discussion of the mass dependence  of the various 
terms of the heavy quark expansion, such 
a calculation has a variety of phenomenological applications, 
of which the most prominent one is its 
application to inclusive semileptonic $b \to c$ transitions.  

These decays are currently believed to be the most precise method 
to determine the CKM matrix element 
$V_{cb}$. In this method, $V_{cb}$ is extracted form the heavy 
quark expansion for the total rate, while the 
heavy quark expansion parameters $\mu_\pi$, $\rho_D$ etc. 
are extracted from the moments of the differential rates. 
Based on this methodology, the theoretical uncertainty 
in $V_{cb}$ has been reduced to a level below 1\%, 
while the total uncertainty (including the experimental 
as well as the uncertainty in the extraction of the 
heavy quark expansion parameters) is at the level of 2\%. 
The current extractions of $V_{cb}$ do not yet include 
the $\alpha_s \mu_G^2$ contributions, which are parametrically 
the largest missing pieces in the analysis.  

From the experimental side, the lepton energies cannot be 
measured to arbitrarily low values. Thus either an 
extrapolation is necessary or one has to include a cut into 
the theoretical predictions. Since an extrapolation 
involves a model dependence, it is more favorable to include 
a lepton-energy cut into the theoretical prediction. 

However, unlike in the numerical study 
of~\cite{Alberti:2013kxa,Gambino:2013rza}, 
such a cut cannot be implemented in an analytical 
calculation, at least not exactly. Thus in order 
to make phenomenological use of the analytical calculation one 
needs to take into account effects of such a cut, 
which needs further study. We plan to return to this in a separate 
publication.

\vskip 0.2cm
{\bf Acknowledgments}\\
\noindent
We acknowledge  
discussions with A.G.~Grozin, B.O.~Lange, J.~Heinonen, and 
T.~Huber.
We thank P.~Gambino for communication on their ongoing work on
radiative corrections to inclusive weak decays.
This work is supported by  
DFG Research Unit FOR 1873 ``Quark Flavors Physics
and Effective Theories''.

\newpage 
\begin{appendices}
\appendix
\section{Master integrals} \label{App:AppendixA}
% the \\ insures the section title is centered below the phrase: AppendixA

Here we present the results for master integrals 
entering our calculation in dimensional regularization with 
$D=4-2\eps$.
The general integral
\ba
S(a,b;m^2,p^2)=\frac{1}{i\pi^{D/2}}\int \frac{d^D k}{(m^2-k^2)^a(-(p-k)^2)^b}
\ea
develops a cut at $p^2=s>m^2$ with a discontinuity
$\rho(a,b;m^2,s)$
that is
\ba
\rho(a,b;m^2,s)=\frac{1}{2\pi i}
\left( S(a,b;m^2,s+i0)-S(a,b;m^2,s-i0)\right)\, .
\ea
It is 
a spectrum of a general sunset 
diagram~\cite{we-annals}
\ba
&&\rho(a,b;m^2,s)=\frac{\Gamma(D/2-b)}{\Gamma(a)\Gamma(b)
\Gamma(D-a-2b+1)}   \nonumber \\
&&z^{D-a-2b}s^{D/2-a-b}
{}_2F_1(D/2-b,1-b;D-a-2b+1;z)
\ea
with $z=1-m^2/s$. Here ${}_2F_1(a,b;c;z)$ is a hypergeometric
function.
In our case $m=m_c$ and $s=m_b^2$.

\subsection{Master integrals at LO: two loop}
\begin{figure}[h!]
\begin{center}
\includegraphics[width=0.25\textwidth]{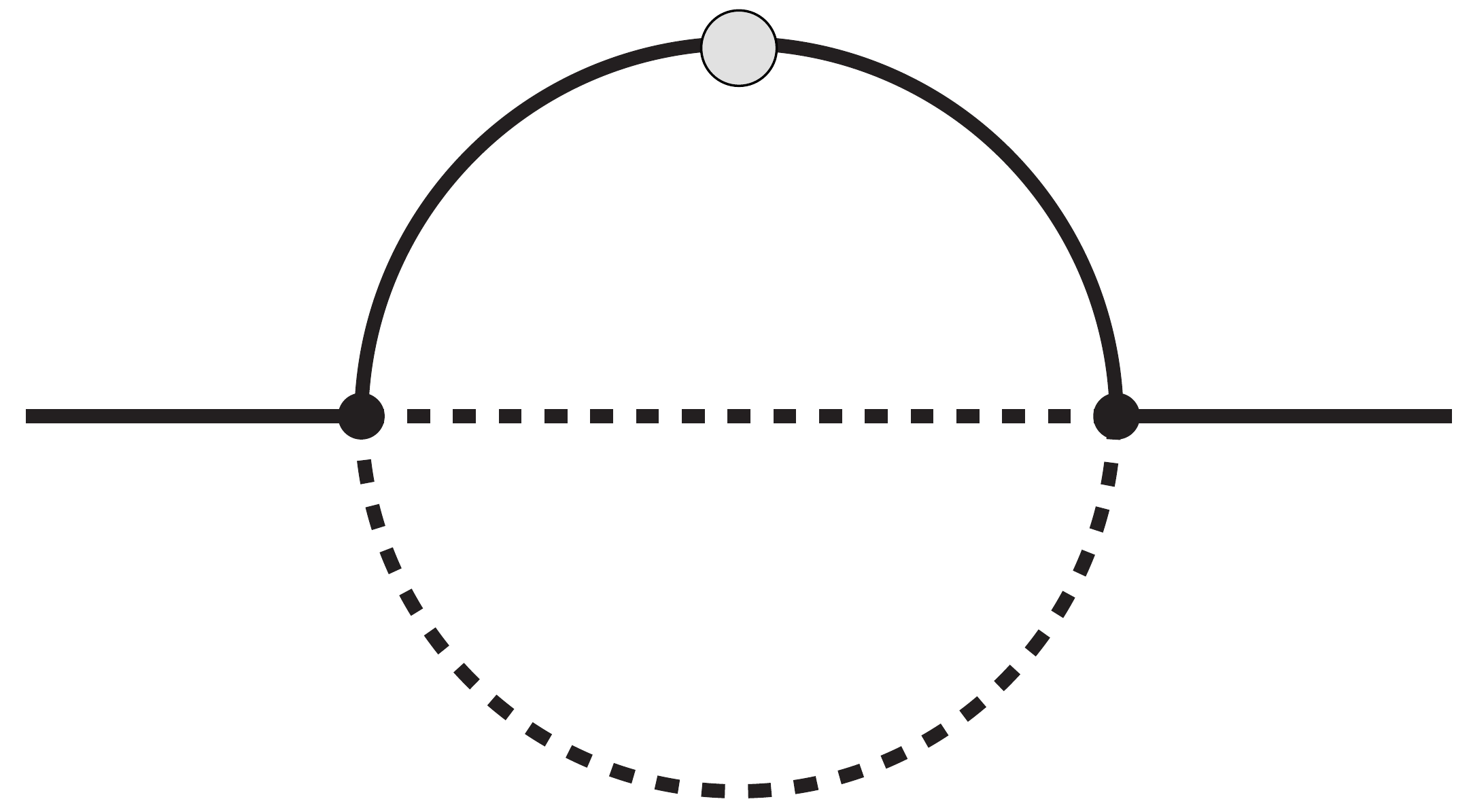}
\end{center}
\caption{\label{fig:mas-LO-twoloop}
Two-loop master integral. A 
dotted line indicates one additional power of the propagator.}
\end{figure}

At LO there are two master integrals.
In both cases it is 
a two loop sunset with one heavy ($m_c$) line.
The internal 
massive line can be a normal one or doubled 
which is denoted by a dot on it, see 
Fig.~\ref{fig:mas-LO-twoloop}.

The closed form for a master integral with a normal line
is
\begin{equation}
M_{00}=S(1,1;0,-1)\rho(1,2-D/2;m_c^2,m_b^2)
\end{equation} 
where $S(1,1;0,-1)$ is a scalar massless loop that is expressible
through $\Gamma$-functions
\begin{equation}
S(a,b;0,-1)=\frac{\Gamma(a+b-D/2)\Gamma(D/2-a)\Gamma(D/2-b)}
{\Gamma(a)\Gamma(b)\Gamma(D-a-b)}
\end{equation} 
We usually set $m_b=1$ in the computation.
The $\eps$-expansion of this integral 
can be obtained with the program HypExp or independently.
At the leading order 
of $\eps$-expansion one has 
\begin{equation}
M_{00}=\frac{1}{2}+m^2 \ln \left(m^2\right)-\frac{m^4}{2}+O(\eps)
\end{equation} 
with $m=m_c$ and $m_b=1$.

The second master integral (dotted) belongs to the same class of sunsets
and can be obtained as a
derivative in $m_c$
\begin{equation}
M_{01}=-\frac{d}{dm_c^2}M_{00}\, .
\end{equation} 
The closed form for this dotted leading order 
master integral is
\begin{equation}
M_{01}=S\left(1,1;0,-1\right)\rho(2,2-D/2;m_c^2,m_b^2)\, .
\end{equation} 

\subsection{Master integrals at NLO: three loop}
At NLO there are master integrals 
that are factorizable, of sunset-type, and nontrivial.

\subsubsection{Factorizable integrals}
The factorized master integrals contain a closed massive loop that can
be either of charmed quark or bottom quark.
\begin{figure}[h!]
\begin{center}
\subfigure[]
{\includegraphics[width=0.25\textwidth]{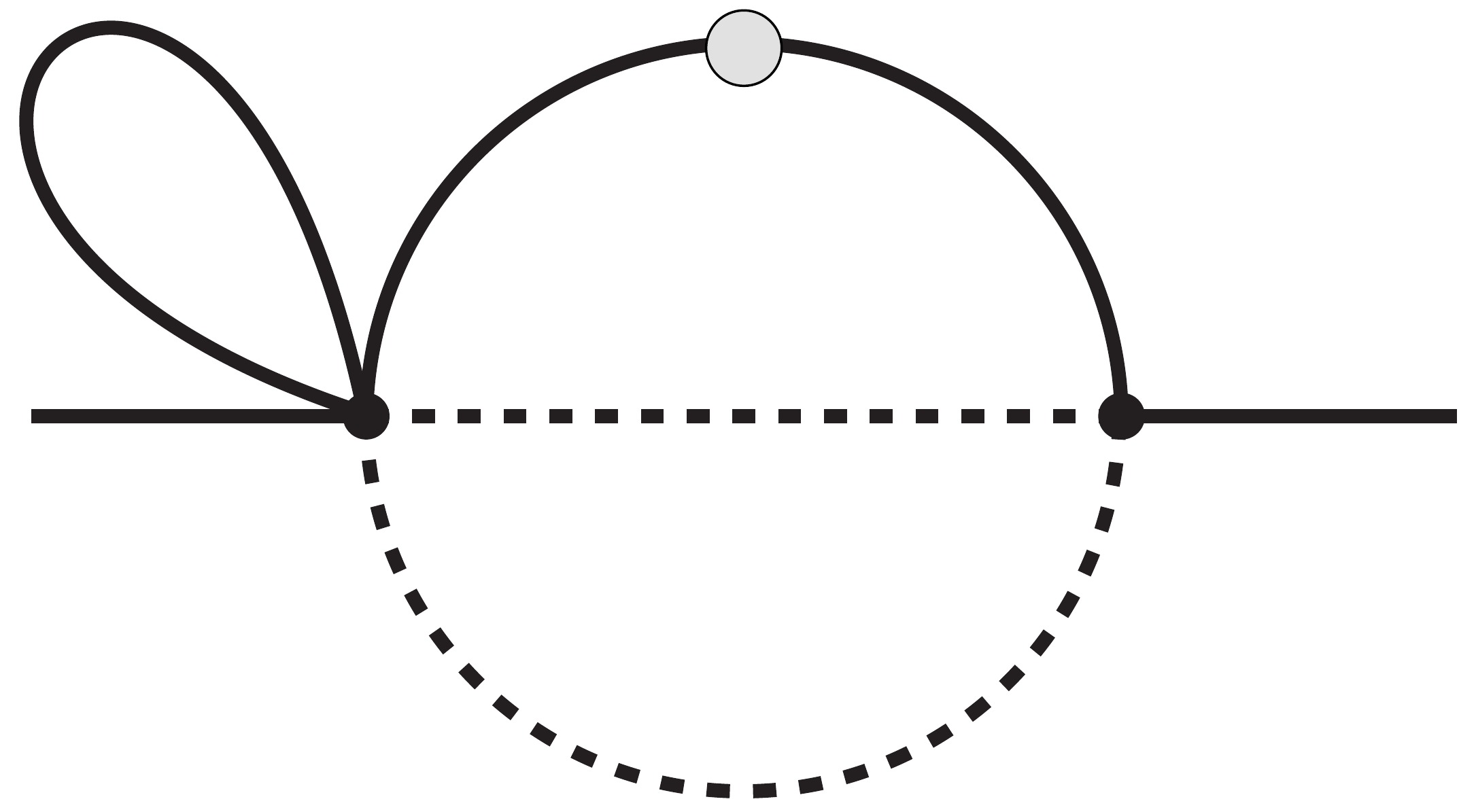}}
%\subfigure[]{\includegraphics[width=0.25\textwidth]{Pictures_masters/M21_22.pdf}}
\end{center}
\caption{\label{fig:mas_fact_threeloop}Factorizable three-loop 
master integrals.}
\end{figure}

These master integrals are:
\begin{equation}
M_{11}=T_0(m_c) M_{00}
\end{equation} 
with $T_0(m)$ being a massive tadpole 
\begin{equation}
T_0(m)=m^{D-2}\Gamma(1-D/2)
\end{equation} 
and the other one 
\begin{equation}
M_{12}=T_0(m_c) M_{01}
\, .
\end{equation} 
The master integrals with a $b$-quark tadpole are
\begin{equation}
M_{41}=T_0(m_b)M_{00}
\end{equation} 
and the other one 
\begin{equation}
M_{42}=T(m_b) M_{01}
\, .
\end{equation} 
In actual computation the bottom quark mass is set to unity.

\subsubsection{Sunset-type integrals}
Non-factorizable but still simple master integrals
$M_{21},M_{22}$
are of the sunset type.
\begin{figure}[h!]
\begin{center}
\includegraphics[width=0.25\textwidth]{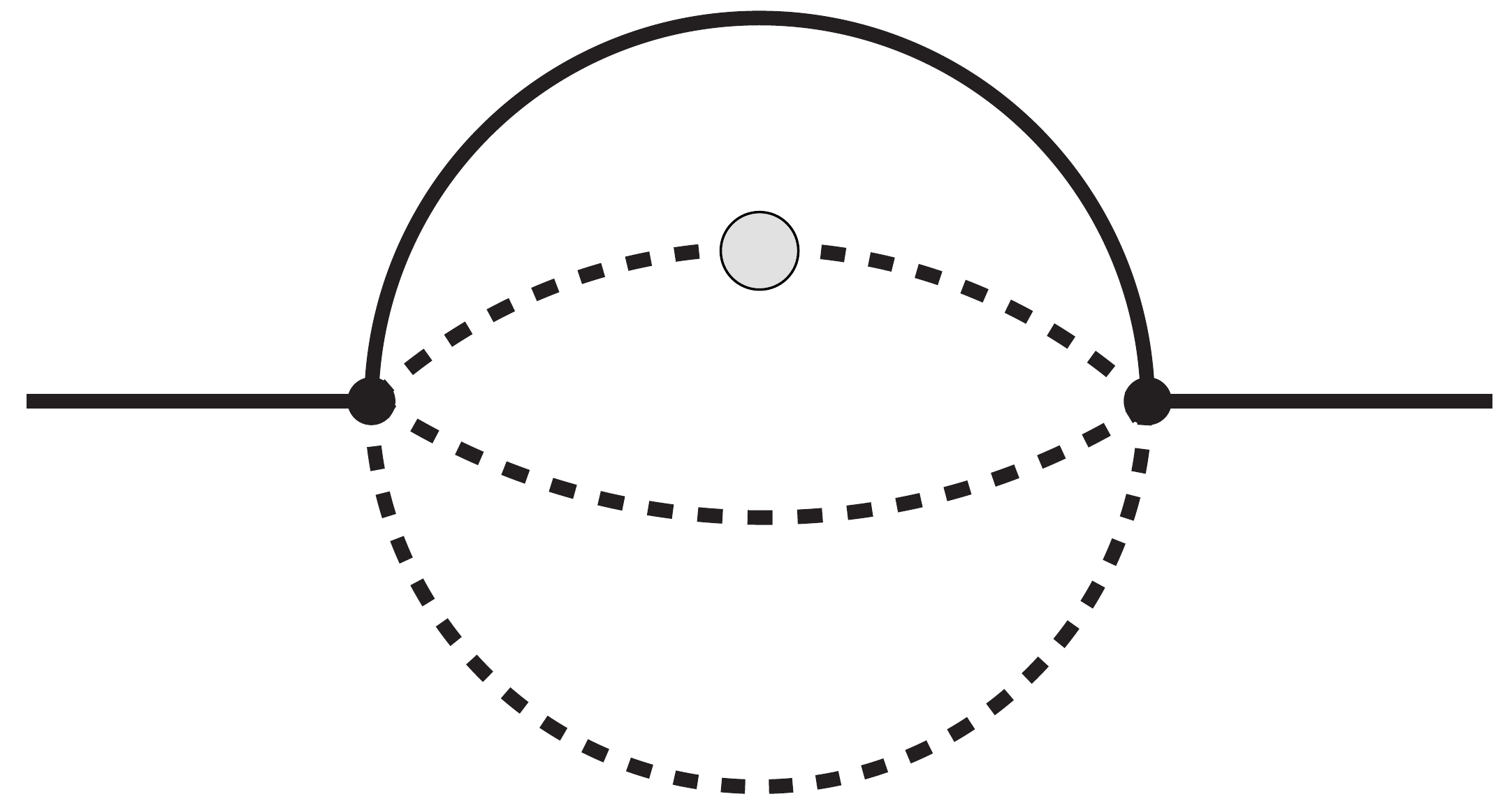}
\end{center}
\caption{\label{fig:mas-sunset_threeloop}
Sunset-type three-loop master integrals.}
\end{figure}

The normal one is given by the basic integral
\begin{equation}
M_{21}=S(1,1;0,-1)S(1,2-D/2;0,-1) \rho(1,3-D,m_c^2,m_b^2)
\end{equation} 
The dotted one is its derivative in loop (charmed quark) 
mass
\begin{equation}
M_{22n}=-\frac{d}{dm_c^2} M_{21}
\end{equation} 
which is again a three loop sunset
\begin{equation}
M_{21}=S(1,1;0,-1)S(1,2-D/2;0,-1) \rho(2,3-D,m_c^2,m_b^2)
\, .
\end{equation} 

\subsubsection{Nontrivial master integrals}
There two nontrivial master integrals that can be chosen in a varity
of ways.
We define the first nontrivial master integral $N_p$
as a sum of left (dotted at bottom line) and
right (dotted on charm line)
diagrams in Fig.~\ref{fig:mas-nontrivial_threeloop_Nmbc}. 
In words, this can be expressed as $N_p=dot.m_b + dot.m_c$. 
\begin{figure}[h!]
\centering 
\includegraphics[width=0.25\textwidth]{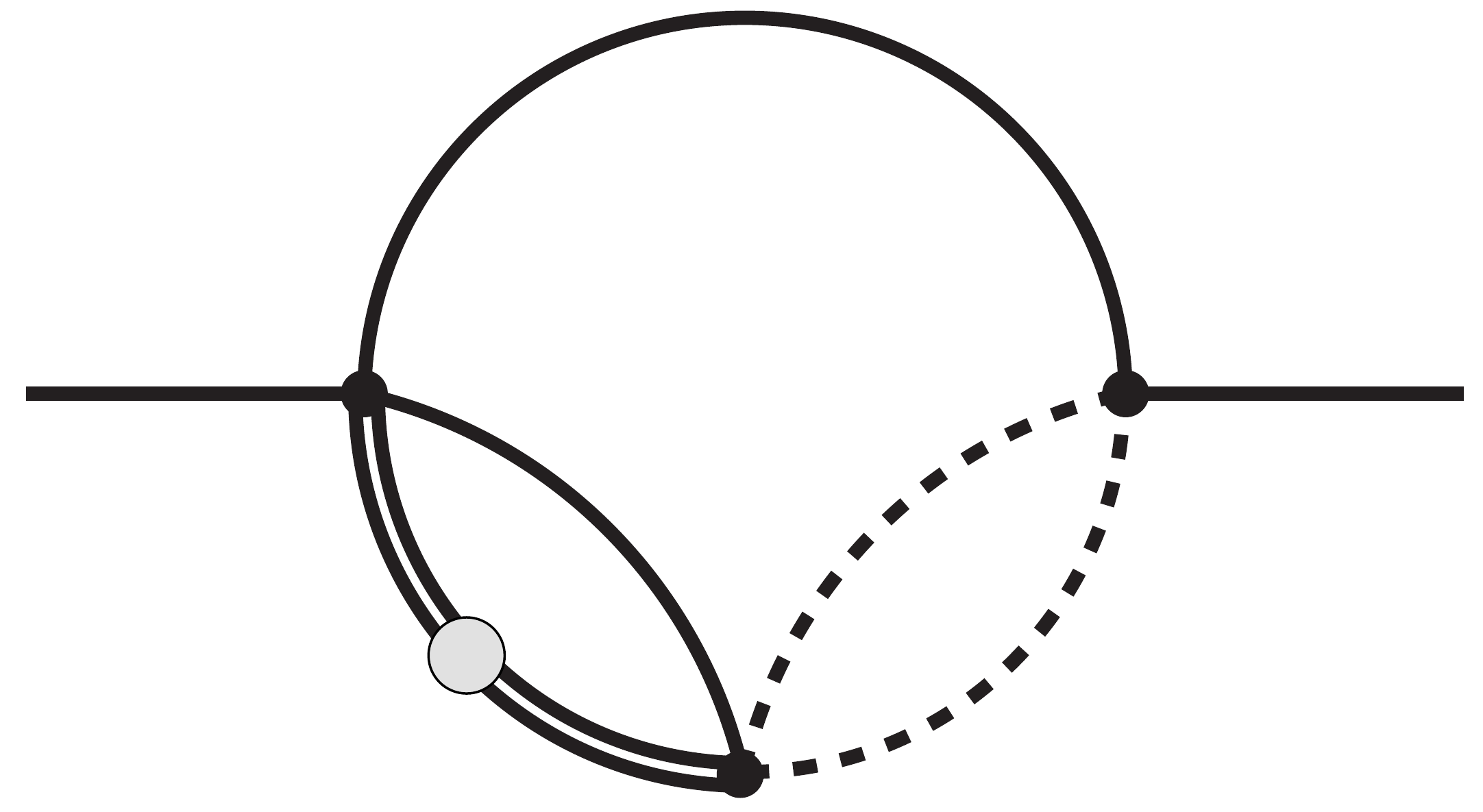}
\qquad
\includegraphics[width=0.25\textwidth]{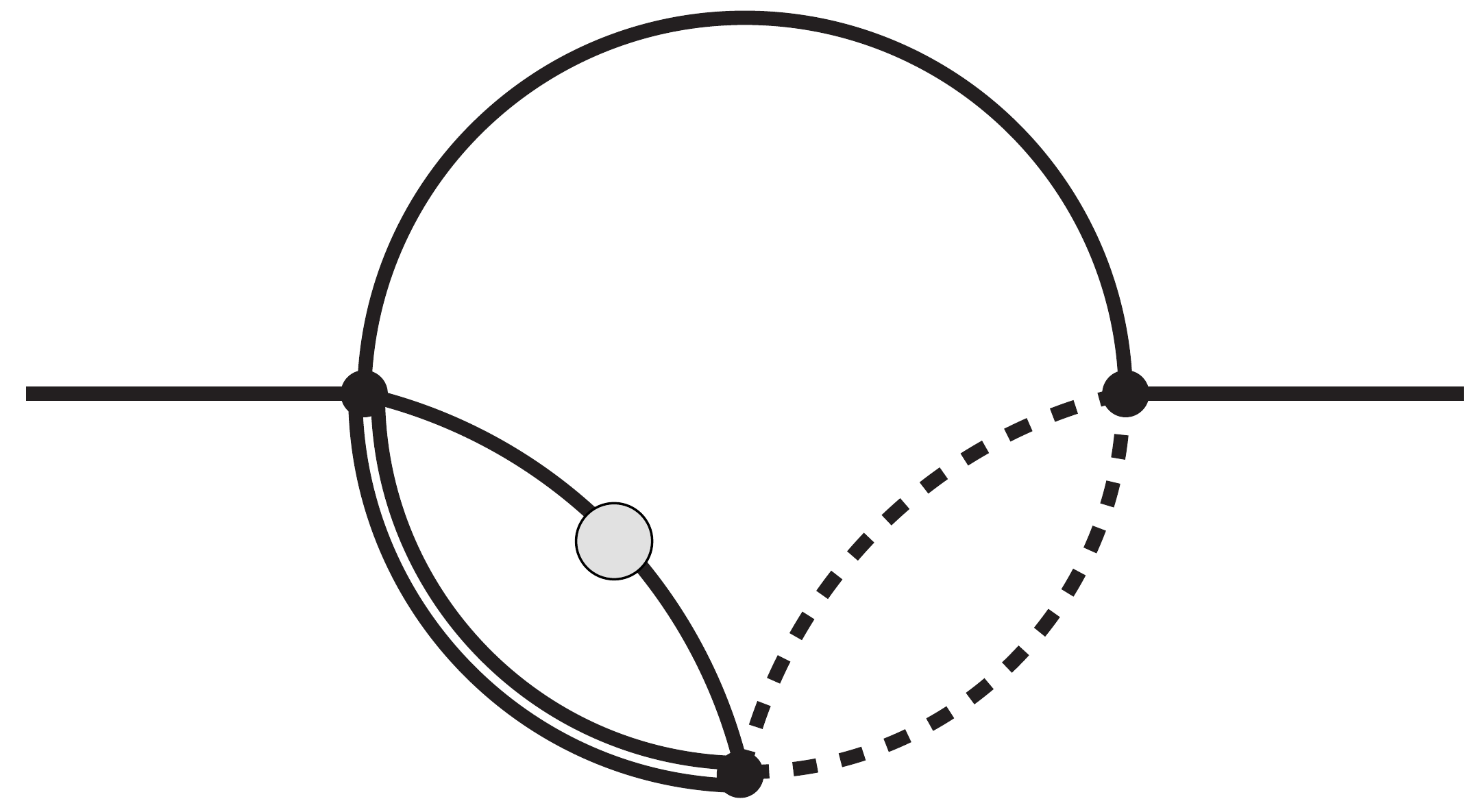}6
\caption{\label{fig:mas-nontrivial_threeloop_Nm6bc}
Nontrivial three-loop masters.}
\end{figure}

We managed to compute the $\eps$-expansion of $N_p$ 
up to the necessary order.
It reads
\ba
N_p=N_p^{LO}+\eps N_p^{NLO}
\ea
with 
\ba
N_p^{LO}=-2(1-r)-(1+r) \log (r)
\ea
and 
\ba
N_p^{NLO}&=&4\sqrt{r} \left(4 (\text{Li}_2\left(-\sqrt{r}\right)-
   \text{Li}_2\left(\sqrt{r}\right))+ \pi ^2
+2\ln (r) \ln \frac{\sqrt{r}+1}{1-\sqrt{r}}\right)
\nonumber \\
&&+\left(\frac{1}{2}-\frac{r}{2}\right)
   \ln ^2(r)
-3\left(r+1\right) \ln (r)
\nonumber \\
&&+4 (r+1) \ln (1-r) \ln (r) + 8(1-r) \ln (1-r)+14 (r-1)
\, .
\ea 
These are master integrals entering partonic contribution for the
total width
and $C_v$ coefficient.

At NLO one more master integral appears to be necessary 
for the $C_G$ coefficient.
It is represented by the difference of left and
right diagrams in Fig.~\ref{fig:mas-nontrivial_threeloop_Nmbc}. 
In words, this can be expressed as $N_p=dot.m_b - dot.m_c$.
We need only the leading term of its $\eps$-expansion that reads
\ba
N_m = (1-r) \left(-4\text{Li}_2(r)+\frac{2 \pi ^2 }{3}
-4 \ln(1-r) \ln (r)+2\ln (r)\right)
-2 r\ln ^2(r)
\, .
\ea

\subsection{Master integrals in massless case}
\begin{figure}[h!]
\begin{center}
\subfigure[]{\includegraphics[width=0.25\textwidth]{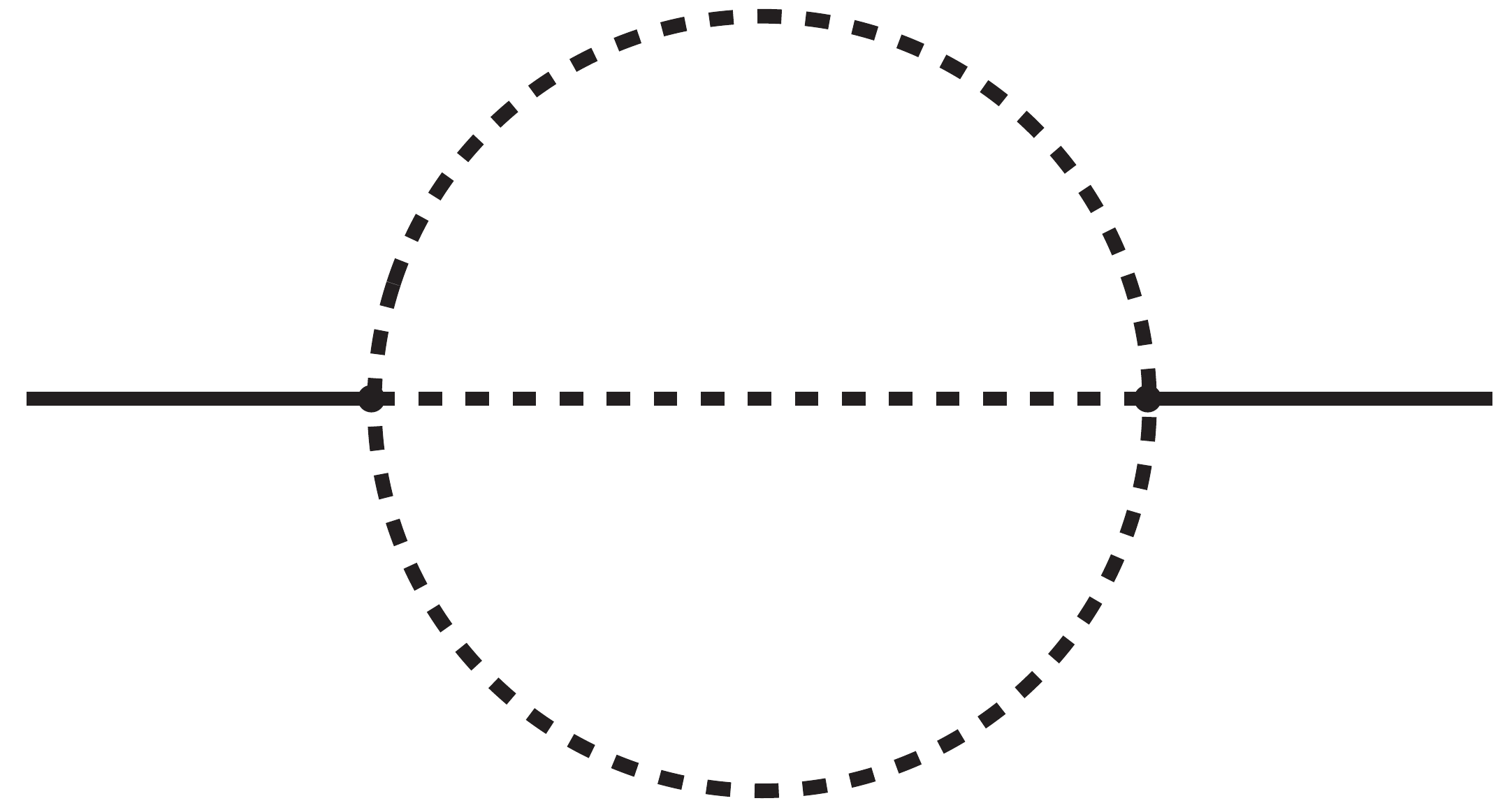}}
%\subfigure[]{\includegraphics[width=0.38\textwidth]{Pictures_masters/m11_massless.pdf}}
\\
\subfigure[]{\includegraphics[width=0.25\textwidth]{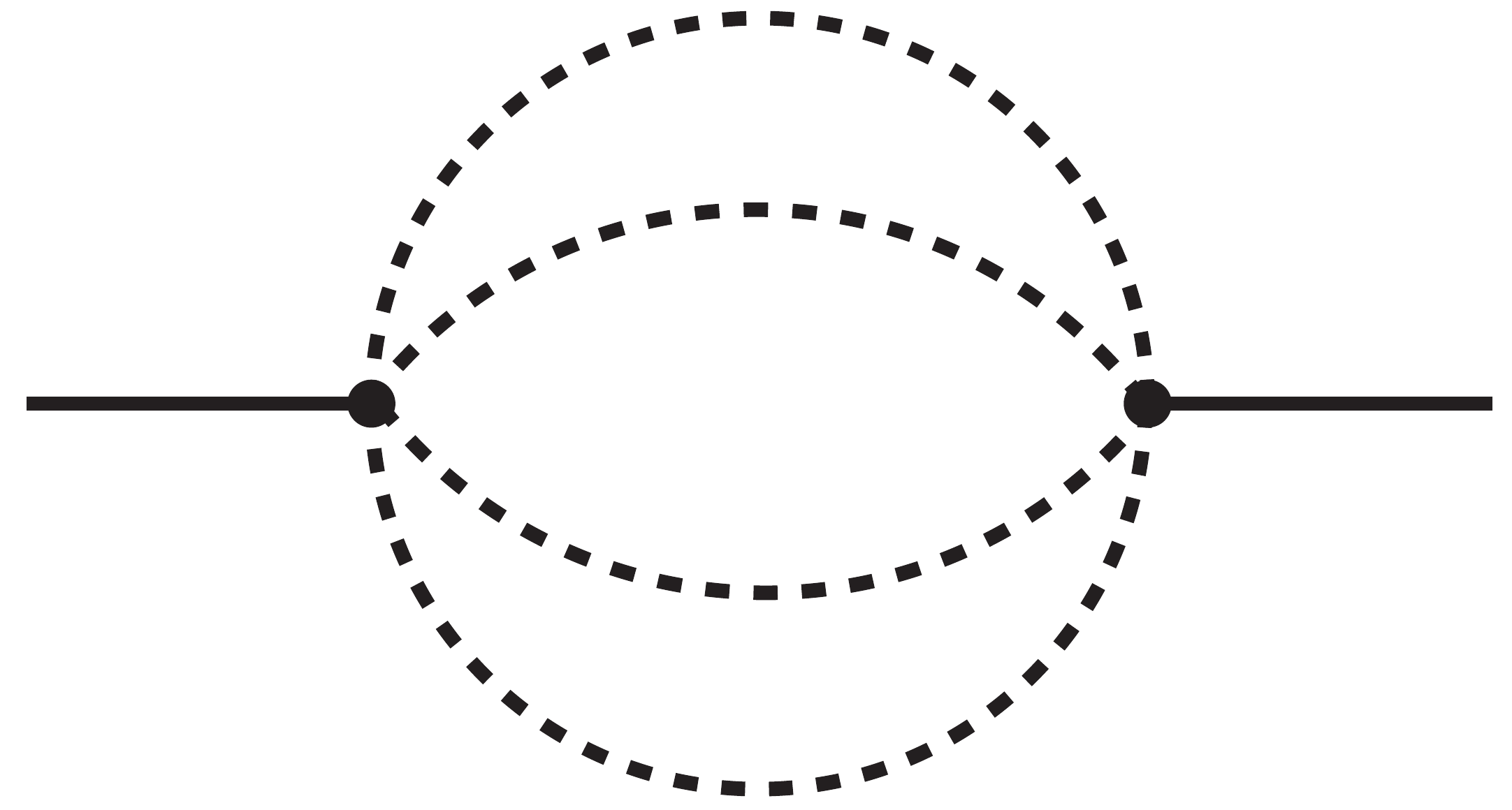}}
\subfigure[]{\includegraphics[width=0.25\textwidth]{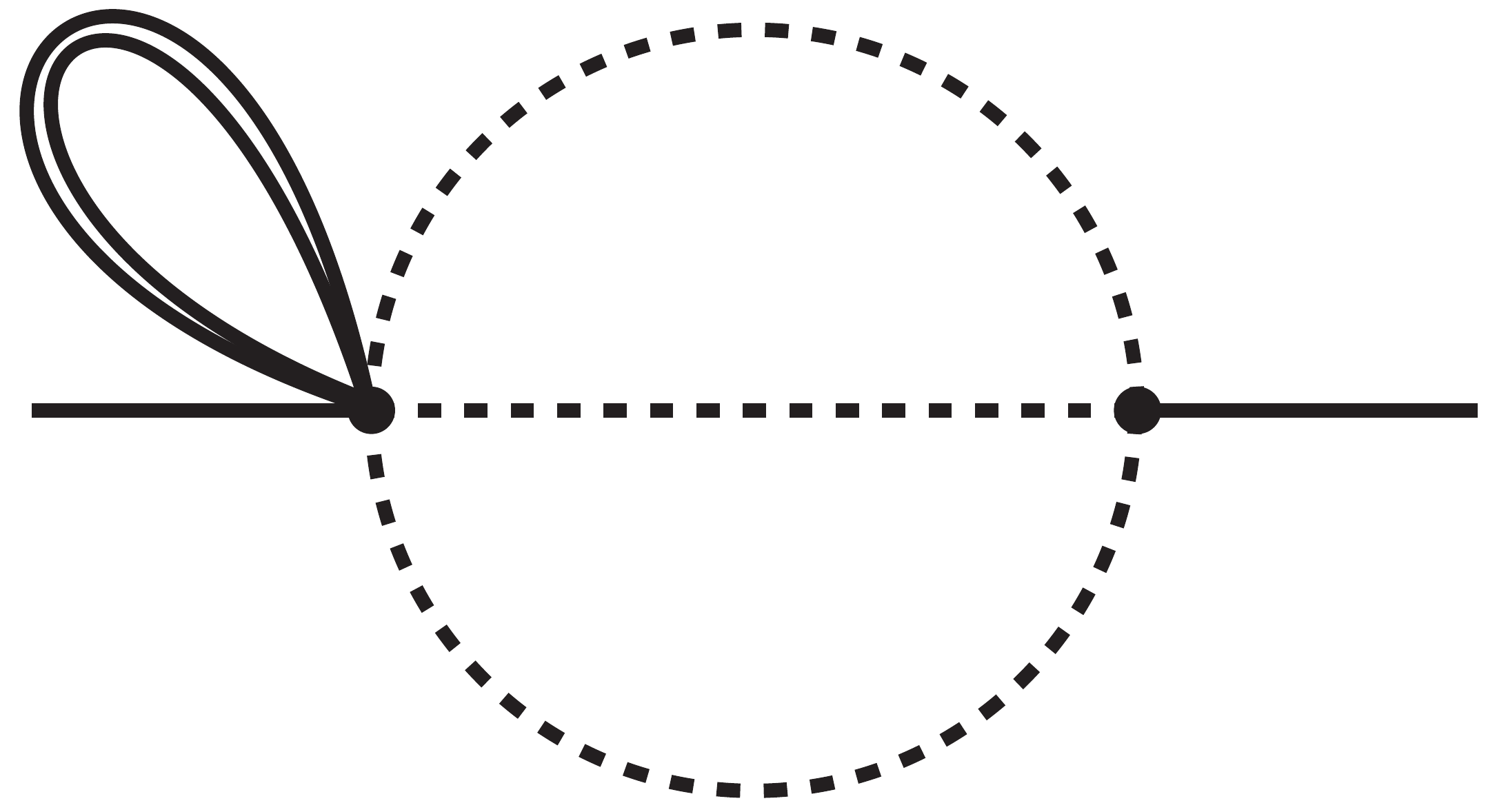}}
\subfigure[]{\includegraphics[width=0.25\textwidth]{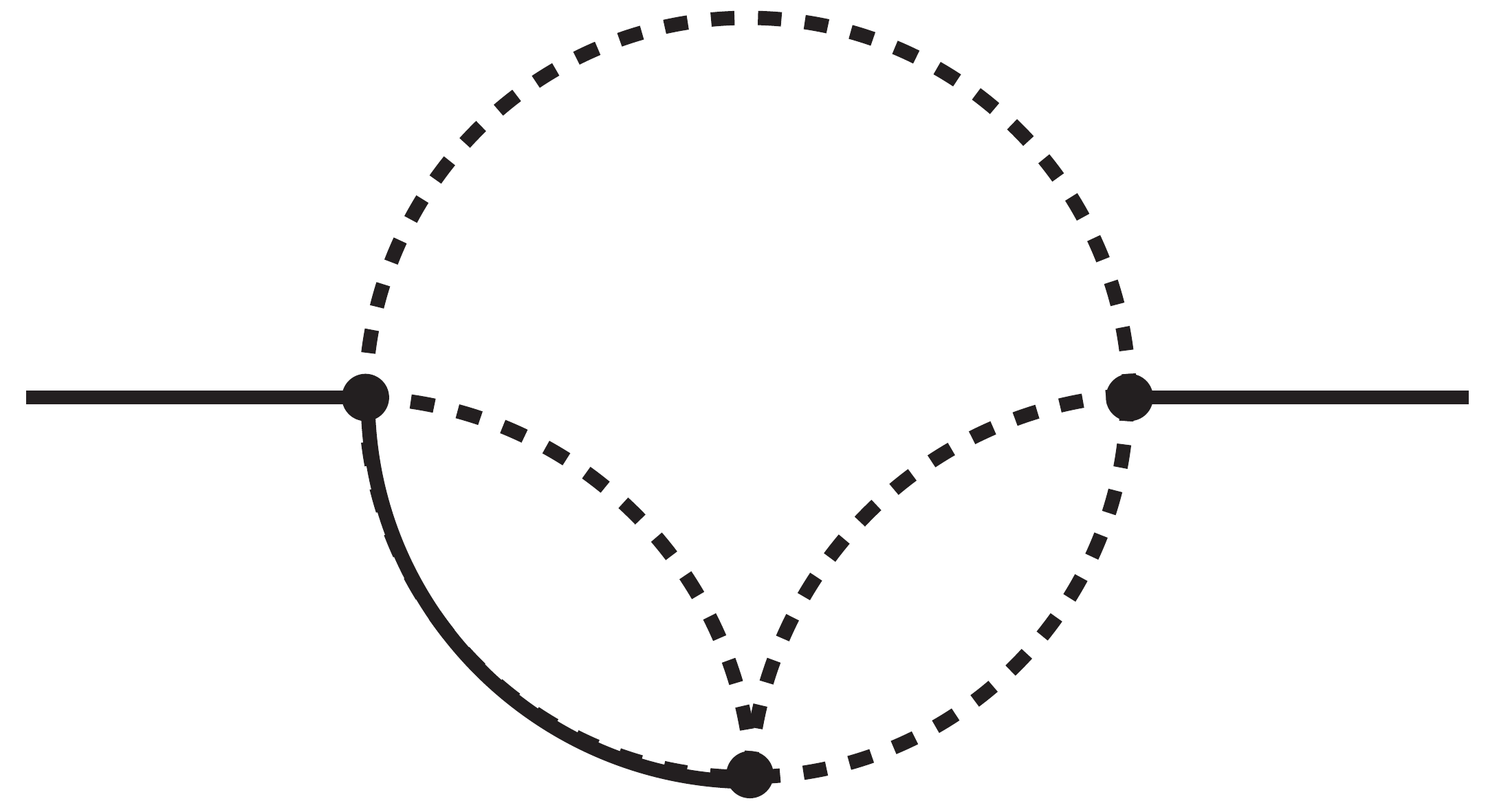}}\\
\end{center}
\caption{\label{fig:mas-two_threeloop_massless}
Massless two- and three-loop master integrals.}
\end{figure}

We have calculated all quantities in the massless limit independently.
The reduction procedure and master integrals 
have been obtained independently as
well.
In massless case master integrals can be found in a concise form.
These master integrals are represented by Feynman diagrams 
given in Fig.~\ref{fig:mas-two_threeloop_massless}.

At leading order there is one master integral
\begin{equation}
M_{00}^0=S(1,1;0,-1)S(1,2-D/2;0,-1) \frac{\sin(2\pi\eps)}{\pi}
\, .
\end{equation} 

At NLO 
in massless case there are three master integrals:

a) factorizable integral
\begin{equation}
M_{11}^0=T_0(m_b) M_{00}^0
\, ;
\end{equation} 

b) sunset integral
\begin{equation}
M_{21}^0=S(1,1;0,-1)S(1,2-D/2;0,-1)S(1,3-D;0,-1)  
\frac{\sin(3\pi\eps)}{\pi}
\, ;
\end{equation} 

c) complicated integral
\begin{equation}
N^0=-S(1,1;0,-1)\frac{\Gamma(1-\eps)^2}{\Gamma(2-\eps)\Gamma(3-3\eps)}
{}_3F_2(\{\eps,1-\eps,1\};\{3-3\eps,2-\eps\},1)
\, .
\end{equation} 

\section{% Appendix  B: 
$C_v$ coefficient at NLO with full mass dependence
} \label{App:AppendixB}
% the \\ insures the section title is centered below the phrase: Appendix B
%
The expression for the coefficient $C_v$ is 
\ba
C_v^{NLO}=&&(3\text{Li}_2(r)-\frac{1}{2} \pi ^2) \left(1-16 r^2-3
   r^4\right)-\frac{1}{24} (1-r) \left(25-1011 r-1487 r^2+189
   r^3\right)
\nonumber\\
&&+\frac{1}{6} r \left(12+450 r+4 r^2+45 r^3\right) \ln (r)
-\frac{1}{6} (1-r) \left(11+11 r+83 r^
2-45 r^3\right) \ln (1-r)\nonumber\\
&&
+\frac{3}{2} r^2 \left(4+r^2\right) \ln ^2(r)+2
   \left(1-30 r^2-3 r^4\right) \ln (1-r) \ln (r)
\nonumber\\
&&+8r^{3/2} (1+3 r)\left(4 \text{Li}_2^- 
-\pi ^2 -2 \ln \left(\frac{1+\sqrt{r}}
{1-\sqrt{r}}\right) \ln (r)\right)
\ea
where ${\rm Li}_2^{-}={\rm Li}_2(\sqrt{r})-{\rm Li}_2(-\sqrt{r})$.

\section{
$C_G$ coefficient at NLO with full mass dependence
} \label{App:AppendixC}

Here we give results for the $C_{\bar\mu_G^2}^{NLO}$ coefficient.
At NLO we give both color structures.

The $C_A$ color structure coefficient of $\alpha_s/\pi$ reads
\begin{align}
C_{\bar\mu_G^2}^{NLO,C_A}&=
\frac{1}{108} (1-r)
   \left(156-4081 r-354 r^2-405 r^3\right)\nonumber\\
   &+\frac{1}{9}
(6{\rm Li}_2(r)- \pi ^2) \left(1-6 r+24 r^2-11 r^3\right)
\nonumber\\
   &-\frac{(1-r)}{54 r} 
\left(15+20 r-196 r^2-292 r^3-27 r^4\right)\ln(1-r)
\nonumber\\
   &-\frac{1}{54} r
   \left(786+972 r+131 r^2-27 r^3\right) \ln(r)
-\frac{2}{9} \left(1+9 r-93 r^2+19 r^3\right) \ln(1-r)\ln(r)
\nonumber\\
   &+\frac{1}{9}
   r \left(9-33 r+5 r^2\right) \ln(r)^2
\nonumber\\
   &+\frac{8}{3}
r^{1/2}(1-\frac{11}{3} r)\left(4\text{Li}_2^-
-\pi ^2-2 \ln(r) \ln\left(\frac{1+\sqrt{r}}{1-\sqrt{r}}\right)\right)
\, .
\end{align}

The $C_F$ color structure is 
\begin{align}
C_{\bar\mu_G^2}^{NLO,C_F}&=
-\frac{1}{216} (1-r) \left(321-13747 r+5421 r^2-3807 r^3\right)
\nonumber\\
&+\frac{1}{18}(6{\rm Li}_2(r)
-\pi ^2) \left(5+72 r-72 r^2-88 r^3+45 r^4\right)
\nonumber\\
   &-\frac{(1-r)}{54
   r} \left(12-19 r+917 r^2-1795 r^3+585 r^4\right)\ln(1-r)
\nonumber\\
   &+\frac{1}{54} r \left(1500-330 r+2668 r^2-585 r^3\right) \ln(r)
\nonumber\\
   &+\frac{2}{9} \left(11+54 r-48 r^2-94 r^3+45 r^4\right)
   \ln(1-r)\ln(r)\nonumber\\
   &-\frac{1}{18} r \left(72+60 r-112 r^2+45 r^3\right) \ln(r)^2
\nonumber\\
   &+\frac{32}{3}(1-\frac{4}{3}r)r^{1/2}
\left( 4{\rm Li}_2^-- \pi^2 -2 \ln(r)
   \ln\left(\frac{1+\sqrt{r}}{1-\sqrt{r}}\right)\right)
\, .
\end{align}

\end{appendices}


\begin{thebibliography}{999}
\bibitem{Charles:2015gya} 
  J.~Charles, O.~Deschamps, S.~Descotes-Genon, H.~Lacker, A.~Menzel, S.~Monteil, V.~Niess and J.~Ocariz {\it et al.},
  %``Current status of the Standard Model CKM fit and constraints on $\Delta F=2$ New Physics,''
  Phys.\ Rev.\ D {\bf 91}, no. 7, 073007 (2015)
  [arXiv:1501.05013 [hep-ph]].
  %%CITATION = ARXIV:1501.05013;%%
  %16 citations counted in INSPIRE as of 23 Jun 2015


\bibitem{Forte:2015cia} 
  S.~Forte, A.~Nisati, G.~Passarino, R.~Tenchini, C.~M.~C.~Calame, M.~Chiesa, M.~Cobal and G.~Corcella {\it et al.},
  %``The Standard Model from the LHC to future colliders: a contribution to the Workshop "What Next" of INFN,''
  arXiv:1505.01279 [hep-ph].
  %%CITATION = ARXIV:1505.01279;%%
  %1 citations counted in INSPIRE as of 23 juin 2015


\bibitem{Butler:2013kdw} 
  J.~N.~Butler {\it et al.}  
[Quark Flavor Physics Working Group Collaboration],
  %``Report of the Quark Flavor Physics Working Group,''
  arXiv:1311.1076 [hep-ex].
  %%CITATION = ARXIV:1311.1076;%%
  %3 citations counted in INSPIRE as of 10 May 2014

\bibitem{Bevan:2014iga} 
  A.~J.~Bevan {\it et al.}  [BaBar and Belle Collaborations],
  %``The Physics of the $B$ Factories,''
  Eur.\ Phys.\ J.\ C {\bf 74}, no. 11, 3026 (2014)
  [arXiv:1406.6311 [hep-ex]].
  %%CITATION = ARXIV:1406.6311;%%
  %43 citations counted in INSPIRE as of 25 juin 2015

\bibitem{Kinoshita:1958ru} %\cite{Kinoshita:1958ru}
  T.~Kinoshita and A.~Sirlin,
  %``Radiative corrections to Fermi interactions,''
  Phys.\ Rev.\  {\bf 113}, 1652 (1959).
  %%CITATION = PHRVA,113,1652;%%

\bibitem{Berman:1958ti} 
  S.~M.~Berman,
  %``Radiative corrections to muon and neutron decay,''
  Phys.\ Rev.\  {\bf 112}, 267 (1958).
  %%CITATION = PHRVA,112,267;%%
  %192 citations counted in INSPIRE as of 30 Apr 2014

\bibitem{vanRitbergen:1998yd} 
  T.~van Ritbergen and R.~G.~Stuart,
  %``Complete two loop quantum electrodynamic contributions to the muon lifetime in the Fermi model,''
  Phys.\ Rev.\ Lett.\  {\bf 82}, 488 (1999).
%  [hep-ph/9808283].
  %%CITATION = HEP-PH/9808283;%%
  %181 citations counted in INSPIRE as of 30 Apr 2014

\bibitem{Pivovarov:2001mw} 
  A.~A.~Pivovarov,
  %``Muon anomalous magnetic moment: A Consistency check for the next-to-leading order hadronic contributions,''
  Phys.\ Atom.\ Nucl.\  {\bf 66}, 902 (2003)
  [Yad.\ Fiz.\  {\bf 66}, 934 (2003)]
  [hep-ph/0110248].
  %%CITATION = HEP-PH/0110248;%%
  %33 citations counted in INSPIRE as of 25 Jun 2015

\bibitem{Kuhn:2003pu} 
  J.~H.~Kuhn, A.~I.~Onishchenko, A.~A.~Pivovarov and O.~L.~Veretin,
  %``Heavy mass expansion, light by light scattering and the anomalous magnetic moment of the muon,''
  Phys.\ Rev.\ D {\bf 68}, 033018 (2003)
  [hep-ph/0301151].
  %%CITATION = HEP-PH/0301151;%%
  %28 citations counted in INSPIRE as of 25 juin 2015

\bibitem{Petrov:2013vka} 
  A.~A.~Petrov and D.~V.~Zhuridov,
  %``Lepton flavor-violating transitions in effective field theory and gluonic operators,''
  Phys.\ Rev.\ D {\bf 89}, no. 3, 033005 (2014)
  [arXiv:1308.6561 [hep-ph]].
  %%CITATION = ARXIV:1308.6561;%%
  %8 citations counted in INSPIRE as of 25 juin 2015


\bibitem{Shifman:1984wx} 
  M.~A.~Shifman and M.~B.~Voloshin,
  %``Preasymptotic Effects in Inclusive Weak Decays of Charmed Particles,''
  Sov.\ J.\ Nucl.\ Phys.\  {\bf 41}, 120 (1985).
%  [Yad.\ Fiz.\  {\bf 41}, 187 (1985)].
  %%CITATION = SJNCA,41,120;%%
  %326 citations counted in INSPIRE as of 14 May 2014

\bibitem{Georgi:1990um} 
  H.~Georgi,
  %``An Effective Field Theory for Heavy Quarks at Low-energies,''
  Phys.\ Lett.\ B {\bf 240}, 447 (1990).
  %%CITATION = PHLTA,B240,447;%%
  %1127 citations counted in INSPIRE as of 02 May 2014

\bibitem{Neubert:1993mb} 
  M.~Neubert,
  %``Heavy quark symmetry,''
  Phys.\ Rept.\  {\bf 245}, 259 (1994).
%  [hep-ph/9306320].
  %%CITATION = HEP-PH/9306320;%%
  %1193 citations counted in INSPIRE as of 02 May 2014

\bibitem{Manohar:2000dt} 
  A.~V.~Manohar and M.~B.~Wise,
  %``Heavy quark physics,''
  Camb.\ Monogr.\ Part.\ Phys.\ Nucl.\ Phys.\ Cosmol.\  {\bf 10}, 1 (2000).
  %%CITATION = CMPCE,10,1;%%
  %211 citations counted in INSPIRE as of 02 May 2014

\bibitem{Mannel:2014xza} %\cite{Mannel:2014xza}
  T.~Mannel, A.~A.~Pivovarov and D.~Rosenthal,
  %``Inclusive semileptonic B decays from QCD with NLO accuracy for power suppressed terms,''
  Phys.\ Lett.\ B {\bf 741}, 290 (2015)
  [arXiv:1405.5072 [hep-ph]].
  %%CITATION = ARXIV:1405.5072;%%
  %3 citations counted in INSPIRE as of 04 juin 2015

\bibitem{Penin:1998wj} %\cite{Penin:1998wj}
  A.~A.~Penin and A.~A.~Pivovarov,
  %``Next-to-next-to-leading order relation between R(e+ e- ---> b anti-b) and Gamma(sl)(b ---> c l neutrino(l)) and precise determination of |V(cb)|,''
  Phys.\ Lett.\ B {\bf 443}, 264 (1998)
  [hep-ph/9805344].
  %%CITATION = HEP-PH/9805344;%%
  %23 citations counted in INSPIRE as of 16 juin 2015

\bibitem{Bigi:1993fe} 
  I.~I.~Y.~Bigi, M.~A.~Shifman, N.~G.~Uraltsev and A.~I.~Vainshtein,
  %``QCD predictions for lepton spectra in inclusive heavy flavor decays,''
  Phys.\ Rev.\ Lett.\  {\bf 71}, 496 (1993).
%  [hep-ph/9304225].
  %%CITATION = HEP-PH/9304225;%%
  %540 citations counted in INSPIRE as of 02 May 2014


\bibitem{Mannel:1991mc} 
  T.~Mannel, W.~Roberts and Z.~Ryzak,
  %``A Derivation of the heavy quark effective Lagrangian from QCD,''
  Nucl.\ Phys.\ B {\bf 368}, 204 (1992).
  %%CITATION = NUPHA,B368,204;%%
  %191 citations counted in INSPIRE as of 14 Feb 2014

\bibitem{Manohar:1997qy} 
  A.~V.~Manohar,
  %``The HQET / NRQCD Lagrangian to order alpha / m-3,''
  Phys.\ Rev.\ D {\bf 56}, 230 (1997).
%  [hep-ph/9701294].
  %%CITATION = HEP-PH/9701294;%%
  %171 citations counted in INSPIRE as of 30 Apr 2014

\bibitem{Benson:2003kp} 
  D.~Benson, I.~I.~Bigi, T.~Mannel and N.~Uraltsev,
  %``Imprecated, yet impeccable: On the 
%theoretical evaluation of Gamma(B ---> X(c) l nu),''
  Nucl.\ Phys.\ B {\bf 665}, 367 (2003).
%  [hep-ph/0302262].
  %%CITATION = HEP-PH/0302262;%%
  %121 citations counted in INSPIRE as of 14 Feb 2014


\bibitem{HigherOrders} 
  T.~Mannel, S.~Turczyk and N.~Uraltsev,
  %``Higher Order Power Corrections in Inclusive B Decays,''
  JHEP {\bf 1011}, 109 (2010).
%  [arXiv:1009.4622 [hep-ph]].
  %%CITATION = ARXIV:1009.4622;%%



\bibitem{vanRitbergen:1999gs} 
  T.~van Ritbergen,
  %``The Second order QCD contribution to 
%the semileptonic b ---> u decay rate,''
  Phys.\ Lett.\ B {\bf 454}, 353 (1999).
%  [hep-ph/9903226].
  %%CITATION = HEP-PH/9903226;%%
  %93 citations counted in INSPIRE as of 14 Feb 2014

\bibitem{Pak:2008qt} 
  A.~Pak and A.~Czarnecki,
  %``Mass effects in muon and semileptonic b ---> c decays,''
  Phys.\ Rev.\ Lett.\  {\bf 100}, 241807 (2008).
 % [arXiv:0803.0960 [hep-ph]].
  %%CITATION = ARXIV:0803.0960;%%
  %52 citations counted in INSPIRE as of 16 May 2014

\bibitem{Melnikov:2008qs}
  K.~Melnikov,
  %``O(alpha(s)**2) corrections to semileptonic decay b ---> cl anti-nu(l),''
  Phys.\ Lett.\ B {\bf 666}, 336 (2008).
%  [arXiv:0803.0951 [hep-ph]].
  %%CITATION = ARXIV:0803.0951;%%
  %44 citations counted in INSPIRE as of 16 May 2014

\bibitem{Becher:2007tk} 
  T.~Becher, H.~Boos and E.~Lunghi,
  %``Kinetic corrections to $B \to X_{c} \ell \bar{\nu}$ at one loop,''
  JHEP {\bf 0712}, 062 (2007).
%  [arXiv:0708.0855 [hep-ph]].
  %%CITATION = ARXIV:0708.0855;%%
  %25 citations counted in INSPIRE as of 14 Feb 2014

\bibitem{Alberti:2013kxa} 
  A.~Alberti, P.~Gambino and S.~Nandi,
  %``Perturbative corrections to power suppressed effects in semileptonic $B$ decays,''
  JHEP, 1 (2014).
%  [arXiv:1311.7381 [hep-ph]].
  %%CITATION = ARXIV:1311.7381;%%
  %2 citations counted in INSPIRE as of 16 May 2014

\bibitem{Grozin:1997ih} 
  A.~G.~Grozin and M.~Neubert,
  %``Higher order estimates of the chromomagnetic moment of a heavy quark,''
  Nucl.\ Phys.\ B {\bf 508}, 311 (1997).
%  [hep-ph/9707318].
  %%CITATION = HEP-PH/9707318;%%
  %19 citations counted in INSPIRE as of 14 Feb 2014

\bibitem{Balk:1993ev} 
  S.~Balk, J.~G.~Korner and D.~Pirjol,
  %``Heavy quark effective theory at large orders in 1/m,''
  Nucl.\ Phys.\ B {\bf 428}, 499 (1994).
%  [hep-ph/9307230].
  %%CITATION = HEP-PH/9307230;%%
  %44 citations counted in INSPIRE as of 30 Apr 2014

\bibitem{Manohar:1993qn} 
  A.~V.~Manohar and M.~B.~Wise,
  %``Inclusive semileptonic B and polarized Lambda(b) decays from QCD,''
  Phys.\ Rev.\ D {\bf 49}, 1310 (1994)
%  [hep-ph/9308246].
  %%CITATION = HEP-PH/9308246;%%
  %522 citations counted in INSPIRE as of 02 May 2014

\bibitem{we-annals}
%\bixbitem{Groote:2005ay} 
  S.~Groote, J.~G.~Korner and A.~A.~Pivovarov,
  %``On the evaluation of a certain class of Feynman diagrams in x-space: Sunrise-type topologies at any loop order,''
  Annals Phys.\  {\bf 322}, 2374 (2007);
%  [hep-ph/0506286].
  %%CITATION = HEP-PH/0506286;%%

  %``A New technique for computing the spectral density of sunset type diagrams: Integral transformation in configuration space,''
  Phys.\ Lett.\ B {\bf 443}, 269 (1998).
%  [hep-ph/9805224].
  %%CITATION = HEP-PH/9805224;%%
  %29 citations counted in INSPIRE as of 30 Apr 2014

\bibitem{reduce}
A.~C.~Hearn, REDUCE, User's manual. Version 3.8.\\
Santa Monica, CA, USA. February 2004

%-Oriented Interactive System for Algebraic Simplification. 

\bibitem{Mathe}Wolfram Research, 
Inc., Mathematica, Version 9.0, Champaign, IL (2012).

\bibitem{FeynCalc}http://www.feyncalc.org/ 

\bibitem{Tkachov:1981wb} 
  F.~V.~Tkachov,
  %``A Theorem on Analytical Calculability of Four Loop Renormalization Group Functions,''
  Phys.\ Lett.\ B {\bf 100}, 65 (1981);\\
  %%CITATION = PHLTA,B100,65;%%
%\bxibitem{Chetyrkin:1981qh} 
  K.~G.~Che\-tyr\-kin and F.~V.~Tkachov,
  %``Integration by Parts: The Algorithm to Calculate beta Functions in 4 Loops,''
  Nucl.\ Phys.\ B {\bf 192}, 159 (1981).
  %%CITATION = NUPHA,B192,159;%%

\bibitem{Lee:2013mka} 
  R.~N.~Lee,
  %``LiteRed 1.4: a powerful tool for the reduction of the multiloop integrals,''
  arXiv:1310.1145 [hep-ph].
  %%CITATION = ARXIV:1310.1145;%%
  %1 citations counted in INSPIRE as of 14 Feb 2014
  
\bibitem{Huber:2007dx} 
  T.~Huber and D.~Maitre,
  %``HypExp 2, Expanding Hypergeometric Functions about Half-Integer Parameters,''
  Comput.\ Phys.\ Commun.\  {\bf 178}, 755 (2008).
%  [arXiv:0708.2443 [hep-ph]].
  %%CITATION = ARXIV:0708.2443;%%
  %78 citations counted in INSPIRE as of 14 Feb 2014

\bibitem{Nir:1989rm} 
  Y.~Nir,
  %``The Mass Ratio m(c) / m(b) in Semileptonic B Decays,''
  Phys.\ Lett.\ B {\bf 221}, 184 (1989).
  %%CITATION = PHLTA,B221,184;%%
  %178 citations counted in INSPIRE as of 02 May 2014

\bibitem{Allison:2008xk} 
  I.~Allison {\it et al.}  [HPQCD Collaboration],
  %``High-Precision Charm-Quark Mass from Current-Current Correlators in Lattice and Continuum QCD,''
  Phys.\ Rev.\ D {\bf 78}, 054513 (2008).
%  [arXiv:0805.2999 [hep-lat]].
  %%CITATION = ARXIV:0805.2999;%%
  %122 citations counted in INSPIRE as of 30 Apr 2014

\bibitem{PDG} 
  J.~Beringer {\it et al.}  [Particle Data Group Collaboration],
  %``Review of Particle Physics (RPP),''
  Phys.\ Rev.\ D {\bf 86}, 010001 (2012).
  %%CITATION = PHRVA,D86,010001;%%
  %3844 citations counted in INSPIRE as of 10 May 2014

\bibitem{Krasnikov:1995is} %\cite{Krasnikov:1995is}
  N.~V.~Krasnikov and A.~A.~Pivovarov,
  %``Running coupling at small momenta, renormalization schemes and renormalons,''
  Phys.\ Atom.\ Nucl.\  {\bf 64}, 1500 (2001)
  [Yad.\ Fiz.\  {\bf 64}, 1576 (2001)];
%  [hep-ph/9510207].
  %%CITATION = HEP-PH/9510207;%%
  %30 citations counted in INSPIRE as of 16 Jun 2015
%\bixxxbitem{Krasnikov:1996jq} 
%  N.~V.~Krasnikov and A.~A.~Pivovarov,
  %``Renormalization schemes and renormalons,''
  Mod.\ Phys.\ Lett.\ A {\bf 11}, 835 (1996).
%  [hep-ph/9602272].
  %%CITATION = HEP-PH/9602272;%%
  %45 citations counted in INSPIRE as of 25 juin 2015


\bibitem{Penin:1998zh} %\cite{Penin:1998zh}
  A.~A.~Penin and A.~A.~Pivovarov,
  %``Next-to-next-to leading order vacuum polarization function of heavy quark near threshold and sum rules for b anti-b system,''
  Phys.\ Lett.\ B {\bf 435}, 413 (1998)
  [hep-ph/9803363].
  %%CITATION = HEP-PH/9803363;%%
  %102 citations counted in INSPIRE as of 16 Jun 2015

\bibitem{Bigi:1994re} 
  I.~I.~Y.~Bigi, A.~G.~Grozin, M.~A.~Shifman, 
N.~G.~Uraltsev and A.~I.~Vainshtein,
  %``On measuring the kinetic energy of the heavy quark inside B mesons,''
  Phys.\ Lett.\ B {\bf 339}, 160 (1994).
%  [hep-ph/9407296].
  %%CITATION = HEP-PH/9407296;%%
  %37 citations counted in INSPIRE as of 30 Apr 2014

\bibitem{Korner:2000xk} %\cite{Korner:2000xk}
  J.~G.~Korner, F.~Krajewski and A.~A.~Pivovarov,
  %``Strong coupling constant from tau decay within renormalization scheme invariant treatment,''
  Phys.\ Rev.\ D {\bf 63}, 036001 (2001)
  [hep-ph/0002166].
  %%CITATION = HEP-PH/0002166;%%
  %39 citations counted in INSPIRE as of 16 juin 2015

\bibitem{Gambino} 
%\bixbitem{Alberti:2012dn} 
  A.~Alberti, T.~Ewerth, P.~Gambino and S.~Nandi,
  %``Kinetic operator effects in B-> X_c l nu at O(alpha_s),''
  Nucl.\ Phys.\ B {\bf 870}, 16 (2013).
%  [arXiv:1212.5082].
  %%CITATION = ARXIV:1212.5082;%%
  %3 citations counted in INSPIRE as of 14 Feb 2014

\bibitem{Falk:1995me}
  A.~F.~Falk, M.~E.~Luke and M.~J.~Savage,
  %``Hadron spectra for semileptonic heavy quark decay,''
  Phys.\ Rev.\ D {\bf 53}, 2491 (1996)
  [hep-ph/9507284].
  %%CITATION = HEP-PH/9507284;%%
  %150 citations counted in INSPIRE as of 24 juin 2015

\bibitem{Gambino:2013rza} %\cite{Gambino:2013rza}
  P.~Gambino and C.~Schwanda,
  %``Inclusive semileptonic fits, heavy quark masses, and V_cb,''
  Phys.\ Rev.\ D {\bf 89}, 014022 (2014).
%  [arXiv:1307.4551 [hep-ph]].
  %%CITATION = ARXIV:1307.4551;%%
  %16 citations counted in INSPIRE as of 16 May 2014



\end{thebibliography}
\end{document}